\documentclass[iop]{emulateapj}
\usepackage{apjfonts}

\begin{document}

\title{GRB 090227B: the missing link between the genuine short and long GRBs}

\author{
M. Muccino\altaffilmark{1,2}, R. Ruffini\altaffilmark{1,2,3}, C.L. Bianco\altaffilmark{1,2}, L. Izzo\altaffilmark{1,2}, A.V. Penacchioni\altaffilmark{1,3}
}

\altaffiltext{1}{Dip. di Fisica and ICRA, Sapienza Universit\`a di Roma, Piazzale Aldo Moro 5, I-00185 Rome, Italy.}
\altaffiltext{2}{ICRANet, Piazza della Repubblica 10, I-65122 Pescara, Italy.}
\altaffiltext{3}{Universite de Nice Sophia Antipolis, Nice, CEDEX 2, Grand Chateau Parc Valrose.}

\shorttitle{GRB 090227B}

\shortauthors{Muccino et al.}

\begin{abstract}

The time-resolved spectral analysis of GRB 090227B, made possible by the \textit{Fermi}-GBM data, allows to identify in this source the missing link between the genuine short and long GRBs.
Within the Fireshell model of the Gamma-Ray Bursts (GRBs) we predict genuine short GRBs: bursts with the same inner engine of the long bursts but endowed with a severely low value of the Baryon load, $B\lesssim 5\times10^{-5}$.
A first energetically predominant emission occurs at the transparency of the $e^+e^-$ plasma, the Proper-GRB (P-GRB), followed by a softer emission, the extended afterglow.
The typical separation between the two emissions is expected to be of the order of $10^{-3}$ -- $10^{-2}$ s.
We identify the P-GRB of GRB 090227B in the first $96$ ms of emission, where a thermal component with the temperature $kT =(517\pm28)$ keV and a flux comparable with the non thermal part of the spectrum is observed. This non thermal component as well as the subsequent emission, where there is no evidence for a thermal spectrum, is identified with the extended afterglow.
We deduce a theoretical cosmological redshift $z = 1.61 \pm 0.14$.
We then derive the total energy $E^{tot}_{e^+e^-} = (2.83\pm0.15)\times10^{53}$ ergs, the Baryon load $B = (4.13\pm0.05)\times10^{-5}$, the Lorentz $\Gamma$ factor at transparency $\Gamma_{tr} = (1.44\pm0.01)\times10^4$, and the intrinsic duration $\Delta t' \sim 0.35$ s. 
We also determine the average density of the CircumBurst Medium (CBM), $\langle n_{CBM}\rangle = (1.90\pm0.20)\times10^{-5}$ particles/cm$^3$. 
There is no evidence of beaming in the system. 
In view of the energetics and of the Baryon load of the source, as well as of the low interstellar medium and of the intrinsic time scale of the signal, we identify the GRB progenitor as a binary neutron star. 
From the recent progress in the theory of neutron stars, we obtain masses of the stars $m_1 = m_2 = 1.34M_\odot$ and their corresponding radii $R_1 = R_2 = 12.24$ km and thickness of their crusts $\sim 0.47$ km, consistent with the above values of the Baryon load, of the energetics and of the time duration of the event.
\end{abstract}


\section{Introduction}\label{sec:0}

The understanding of GRBs is among the most fascinating and profound conceptual problems of relativistic astrophysics.
Observations at high energies from space missions, such as BATSE \citep{Meegan1992}, Beppo-SAX \citep{Metzger1997}, Swift Burst Alert Telescope (BAT) \citep{Gehrels2005}, AGILE \citep{AGILE}, Fermi Gamma-ray Burst Monitor (GBM) \citep{Meegan2009} and others, have revealed that GRBs emit in a few seconds of the time of the observer almost the energy equivalent to a solar mass.
This allows the observability of these sources over the entire visible Universe.

The first systematic analysis on the large sample of GRBs observed by the BATSE instrument on board the Compton Gamma-Ray Observer (CGRO) satellite \citep{Meegan1992}, evidenced a bi-modal temporal distribution in the $T_{90}$ observed duration of prompt emission of GRBs.
The ``long'' and ``short'' GRBs were defined as being longer or shorter than $T_{90} = 2$ s.

Another fundamental progress was achieved by \textit{Beppo-SAX} with the discovery of a prolonged soft X-ray emission, the ``afterglow'' \citep{Costa1997}, following the traditional hard X-ray emission observed by BATSE, that was called the ``prompt emission''.

In recent years, the observations by the \textit{Swift} satellite \citep{Gehrels2005} evidenced the existence of a possible third class of burst presenting hybrid properties between the short and the long ones: the Norris-Bonnell sources \citep{Norris2006}.
The prompt emission of these sources is characterized by an initial short spike-like emission lasting a few seconds, followed by a prolonged softer extended emission lasting up to some hundred seconds.
They were initially indicated in the literature as ``short GRBs with an extended emission''. 

In parallel the theoretical progress in the Fireshell model of GRBs \cite[see][]{Ruffini2001c,Ruffini2001,Ruffini2001a} has led to an alternative explanation of the Norris-Bonnell sources as ``disguised short bursts'' \citep{Bernardini2007,Bernardini2008,Caito2009,Caito2010,deBarros2011}: canonical long bursts exploding in halos of their host galaxies, with $\langle n_{CBM}\rangle \approx 10^{-3}$ particles/cm$^3$ (see Sec.~\ref{sec:fireshell:disguised}).

\begin{figure}
\centering
\includegraphics[width=\hsize,clip]{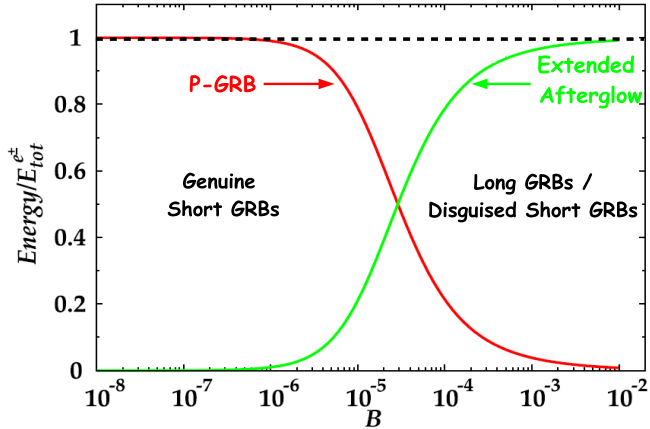}
\caption{The energy emitted in the extended afterglow (solid green curve) and in the P-GRB (solid red curve) in units of $E_{e^+e^-}^{tot} = 1.77 \times 10^{53}$ erg (dashed horizontal line), as functions of $B$. The crossing point, corresponding to the condition $E_{P-GRB} \equiv 50\%E_{e^+e^-}^{tot}$, marks the division between the genuine short and disguised short and long GRBs region.}
\label{fig:1}
\end{figure}

The aim of this article, using the data obtained by the \textit{Fermi}-GBM satellite \citep{Meegan2009}, is to probe the existence of a yet new class of GRBs which we here define ``genuine short GRBs'', theoretically predicted by the Fireshell model \citep{Ruffini2001, Ruffini2002}.
This class of canonical GRBs is characterized by severely small values of the Baryon load, $B \lesssim 10^{-5}$ (see Fig.~\ref{fig:1}).
The energy emitted in the P-GRB is predominant and the characteristic duration is expected to be shorter than a fraction of a second (see Sec.~\ref{sec:fireshell:genuine}).

We have started a search for these genuine short GRBs among the bursts detected by the Fermi GBM instrument, in its first three years of mission. 
The initial list of short GRBs was reduced requiring that no prominent X-rays and optical afterglows be observed. 
Among these bursts we have identified GRB 090227B.
From its observed light curves, we have performed the spectral analysis of the source, and within the theory, we have inferred its cosmological redshift, and all the basic parameters of the burst, as well as the isotropic energy, the Lorentz $\Gamma$ factor at transparency, and the intrinsic duration.

In Sec.~\ref{sec:fireshell} we recall the relevant properties of the Fireshell model.
In Sec.~\ref{sec:analysis} we report the observation of GRB 090227B by the different satellites and the data analysis.
In Sec.~\ref{sec:3} we determine all the parameters characterizing GRB 090227B within the Fireshell scenario, including the redshift.
In Sec.~\ref{sec:4} we provide an estimation of the lower limit on the Lorentz $\Gamma$ factor from the definition of opacity, finding the agreement with the theoretically determined Lorentz $\Gamma$ factor.  
In the conclusions we show that GRB 090227B is the missing link between the genuine short and the long GRBs, with some common characteristics between the two classes.
Further analysis of genuine short GRB with a yet small value of B should lead to P-GRB with a yet more pronounced thermal component.
We identify the progenitor of GRB 090227B as a symmetric binary system of two neutron stars, each of $\sim1.34M_\odot$.

\section{The Fireshell vs the Fireball model and the issue of the photospheric emission}\label{sec:fireshell}

Soon after the announcement of the discovery of GRBs \citep{Strong1975}, \citet{Damour} proposed to explain the energy source of GRBs in terms of the $e^+e^-$ pair plasma created in the process of vacuum polarization during the formation of a Kerr-Newman black hole. They mentioned that the energetics to be expected in this model is approximately $10^{54}$--$10^{55}$ erg for a $10M_\odot$ black hole. At the time nothing was known about the energetics of GRBs, being their distance unknown. They did not pursue further the details of the model pending additional observational evidence.

The idea of the role of an $e^+e^-$ pair plasma as energy source of GRBs was proposed again and independently by \citet{CavalloRees}. They proposed a sudden release of energy in a process of gravitational collapse leading to a large number of $e^+e^-$ pairs , whose instantaneous annihilation would lead to a vast release of energy pushing on the CBM: the concept of ``fireball''.

The concept of fireball was further examined by \citet{Goodman1986}, who quantified the dynamical effects of the expansion of the fireball computing the effect of the blue-shift due to the bulk Lorentz $\Gamma$ factor on the observed temperature.
\citet{ShemiPiran1990} were among the first to compute the dynamics of such a fireball in presence of baryonic matter, described by the adimensional parameter $\eta = E_0/M_Bc^2$, in which $E_0$ is the initial total energy of the fireball. They clearly pointed out that for large values of $\eta$, photons carry most of the energy of the fireball. In the opposite regime most of $E_0$ is converted in the kinetic energy of the baryons and only a small fraction is carried away by the photons at transparency. Further works were presented by \citet{1993ApJ...415..181M}, \citet{PiranShemiNarayan} and \citet{Katz1994a}.

After the discovery by Beppo-SAX \citep{Costa1997} of the cosmological nature of GRBs \citep{vanParadijs1997}, it became clear that the energetics presented by \citet{Damour} was indeed correct and their work represented one of the handful GRB models still viable \citep{RRKl}. The return to the model led to a further step in the comprehension of GRBs \citep{RSWX2,RSWX} with the detailed analysis of the rate equation which accounts for the gradual annihilation of the pairs, in a relativistic expanding shell, during the entire optically thick acceleration phase of GRBs: the concept of ``fireshell''.

The main differences between the fireball and the fireshell scenarios are outlined in the paper of \citet{Bianco2006}, while in \citet{2007PhRvL..99l5003A} it was definitely proved that in an optically thick $e^+e^-$ plasma the annihilation of the pairs does not occur instantaneously, as originally assumed by Cavallo and Rees. Instead the optically thick $e^+e^-$ plasma reaches the thermal equilibrium in a very short time scale, $\sim10^{-12}$ s, and then dynamically expands following the approach in \citet{RSWX2,RSWX}.

In the meantime the BATSE observations led to a phenomenological classification of GRBs, based on their observed duration, into ``long'' and ``short'' GRBs \citep{Klebesadel1992,Dezalay1992,Koveliotou1993,Tavani1998}. Initially this fact was interpreted in terms of different progenitors for these two classes \cite[see][]{Blinnikov1984,Woosley1993,Paczynski1998}.

\begin{figure*}
\centering
\includegraphics[width=0.49\hsize,clip]{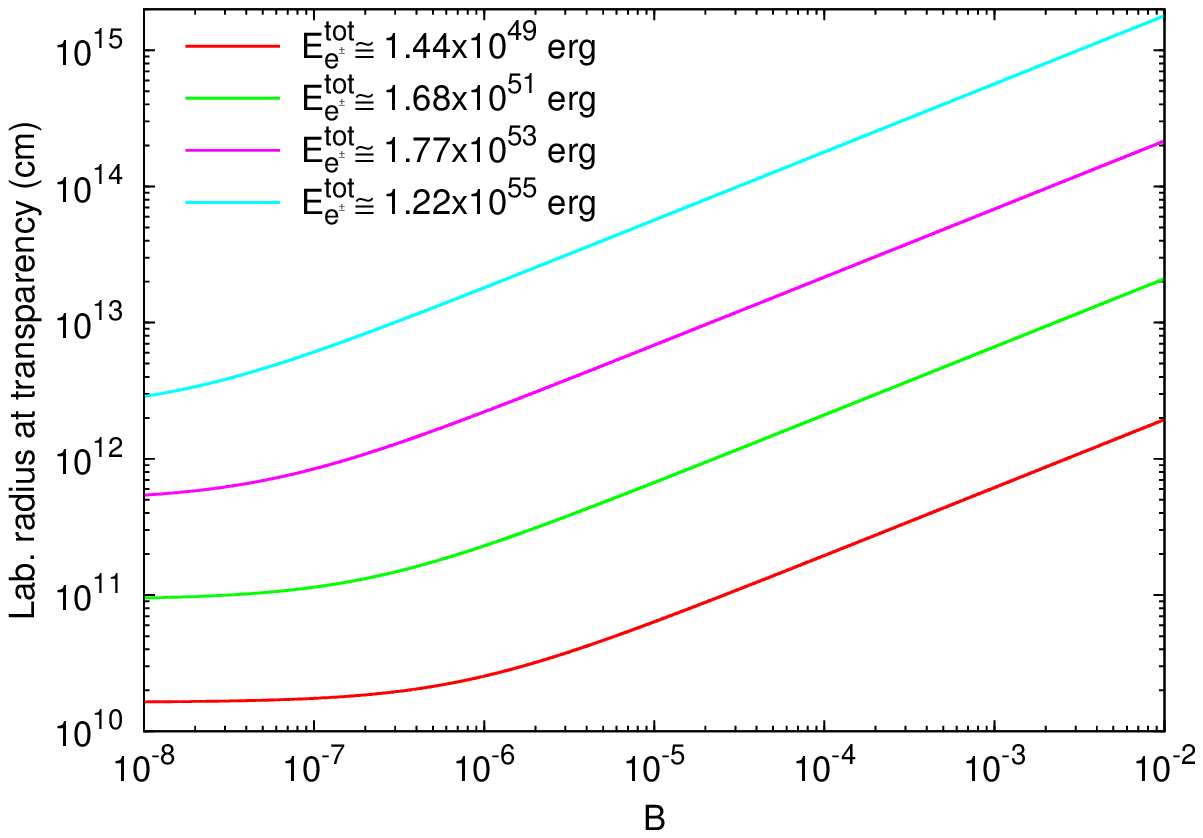}
\includegraphics[width=0.49\hsize,clip]{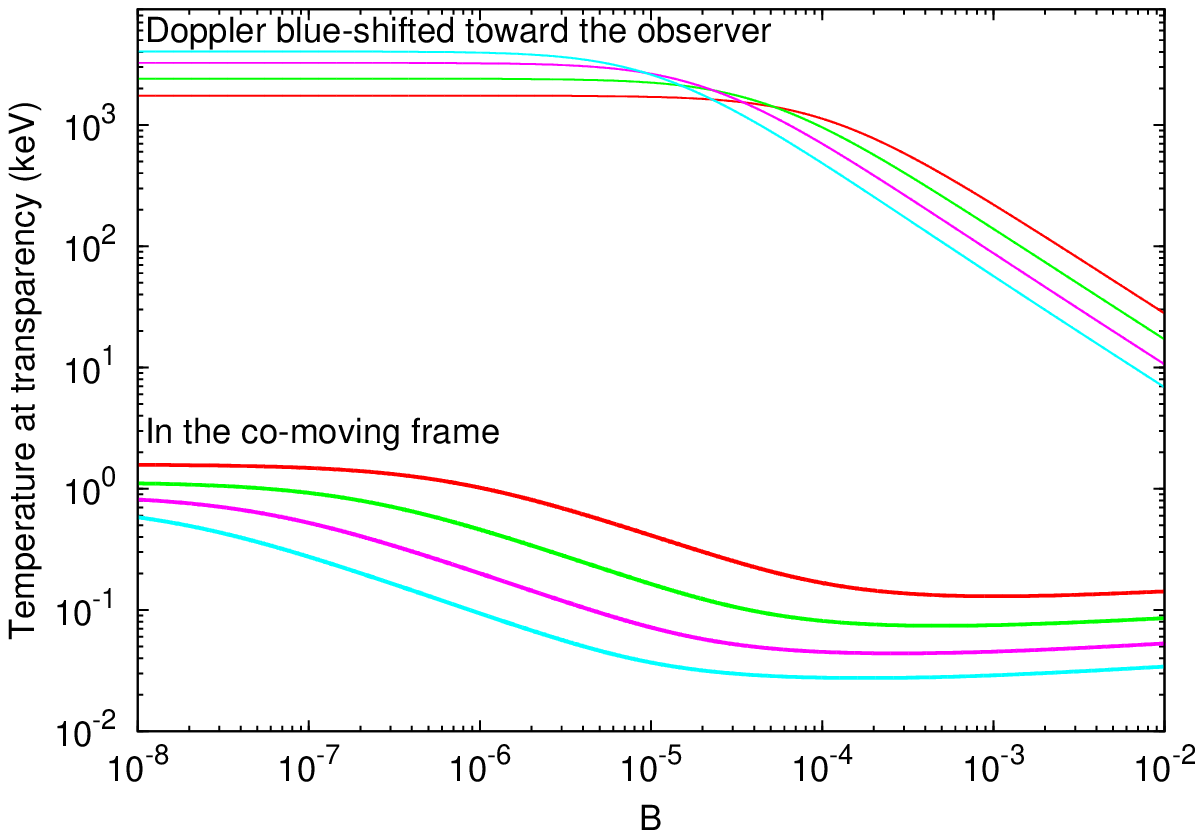}
\includegraphics[width=0.49\hsize,clip]{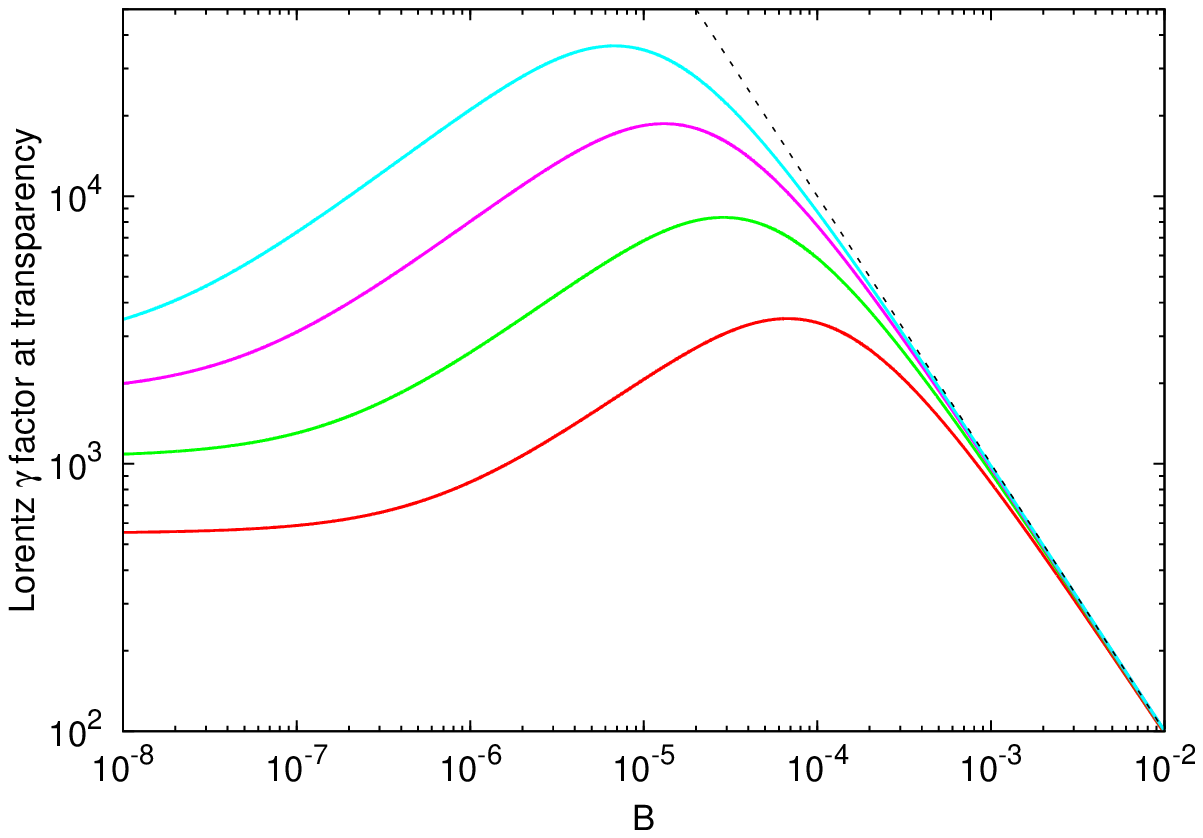}
\includegraphics[width=0.49\hsize,clip]{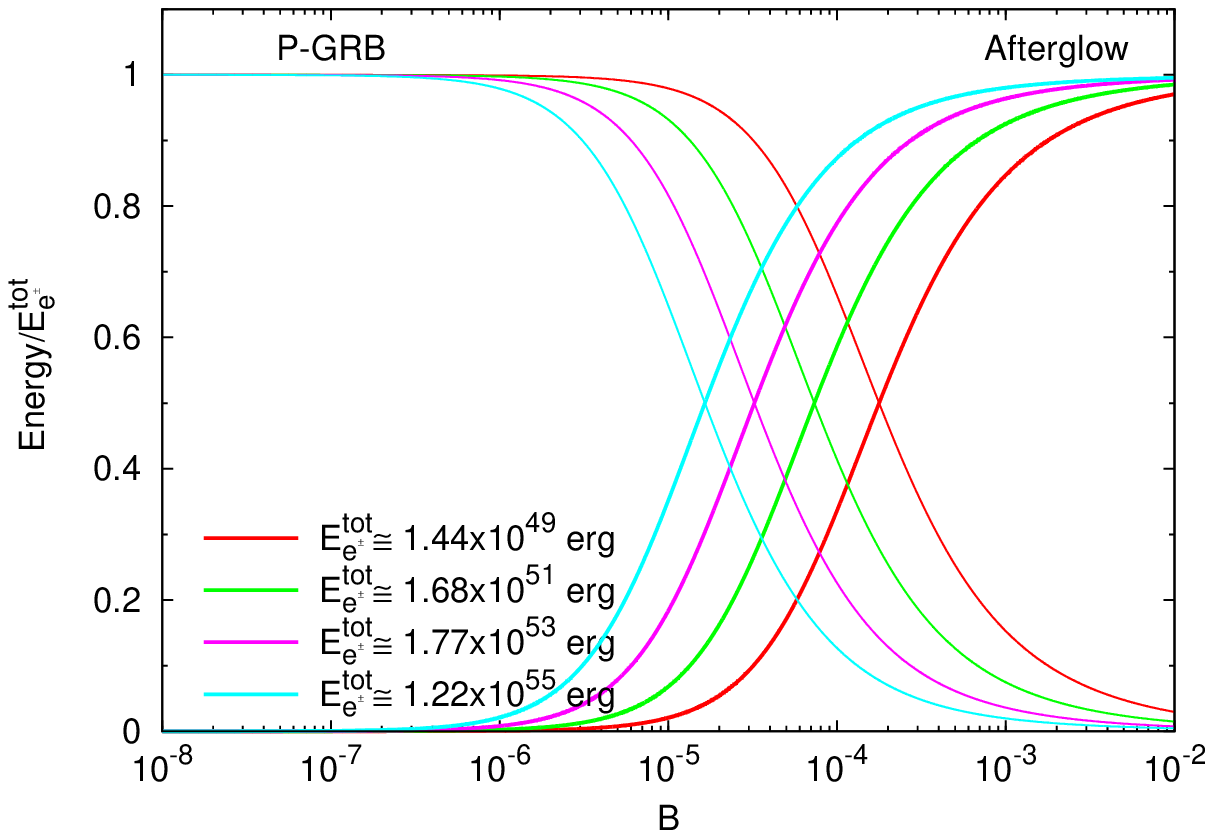}
\caption{The main quantities of the Fireshell model at the transparency for selected values of $E_{e^+e^-}^{tot}$: the radius in the laboratory frame, the co-moving frame and blue-shifted toward the observer temperatures of the plasma, the Lorentz $\Gamma$ factor, and the fraction of energy radiated in the P-GRB and in the extended afterglow as functions of $B$. In these simulations a sudden transition between the optically thick adiabatic phase and the fully radiative condition at the transparency has been assumed.}
\label{fig:4}
\end{figure*}

In 2001 an interpretation within the Fireshell model was proposed to explain the differences between the short and the long GRBs. This interpretation was based on the Baryon load $B$ (inverse of $\eta$). In this picture, both long and short GRBs originate from the same basic machine, the dyadotorus, from an implosion leading to the formation of a Kerr-Newman black hole \citep{RRKerr}. The long bursts correspond to GRBs with $B \gtrsim 3.0\times10^{-4}$ and the short ones to GRBs with $B \lesssim 10^{-5}$ (Fig.~\ref{fig:1}). For $10^{-5} \lesssim B \lesssim 3.0\times 10^{-4}$ it depends also on the value of the total energy of the pairs $E_{e^+e^-}^{tot}$ (see Fig.~\ref{fig:4}). The short bursts should have in the limit of $B \to 0$ no afterglow. This was followed in 2002 by a further theoretical work evidencing also the relevance of an additional parameter influencing the interpretation above classification: the average density of the environment CBM \citep{Ruffini2002,Ruffini2004,Ruffini2005IJMPD}. This led to the new concept of ``disguised short'' GRBs \citep{Bernardini2007,Bernardini2008,Caito2009,Caito2010,deBarros2011}.

Let us briefly go in some more detail in the fireshell model. 
As we have recalled, the GRBs originate from the process of vacuum polarization occurring in the formation of a black hole, resulting in pair creation \citep{Damour,Xue2008,RVX2010}.
The formed $e^+e^-$ plasma, with total energy $E_{e^+e^-}^{tot}$, reaches the thermal equilibrium almost instantaneously \citep{2007PhRvL..99l5003A}. 
The annihilation of these pairs occurs gradually and it is confined in an expanding shell, called \textit{fireshell}, which self-accelerates up to ultra relativistic velocities \citep{RSWX2}, and engulfs the baryonic matter (of mass $M_B$) left over in the process of collapse, which thermalizes with the pairs due to the large optical depth \citep{RSWX}. 
The Baryon load is measured by the dimensionless parameter $B = M_Bc^2/E_{e^+e^-}^{tot}$. 
The fireshell continues to self-accelerate until it reaches the transparency condition and a first flash of radiation, the P-GRB, is emitted \citep{Ruffini2001}. 
The radius at which the transparency occurs and the theoretical temperature, the Lorentz factor as well as the amount of the energy emitted in the P-GRB are functions of $E_{e^+e^-}^{tot}$ and $B$ (see Fig.~\ref{fig:4}).

In recent years a systematic analysis of the possible presence of a thermal component in the early phases of the prompt emission of GRBs has been performed using the earlier data from BATSE all the way to the latest ones from Fermi \citep{Ryde2004,Ryde2009,BandBB}.
The presence of episodes with a significant thermal component lasting typically from $20$ to $50$ s has been evidenced. 
In some specific cases the thermal component has been shown to vary with time following a broken power-law \citep{Ryde2004,Ryde2009}. 
This problematic has led to the study of the so-called photospheric emission \citep{ReesMeszaros2005,Peer2005,Peer2006,Lazzati2010}.
It has been pointed out \citep{TEXAS,Izzo2012b,Izzo2012,Penacchioni2011} that a marked difference exists between these prolonged emissions occurring at $\Gamma\sim1$ and the specific ones of the $e^+e^-$ recombination occurring at ultra relativistic regimes, $\Gamma>10^2$, and lasting at most a few seconds.
In the specific cases of GRB 970828 \citep{Izzo2012b}, GRB 090618 \citep{Izzo2012} and GRB 101023 \citep{Penacchioni2011} the existence of these two components has been evidenced. 
The first component, at $\Gamma\sim1$, has been associated to the Proto Black Hole (PBH), while the one at $\Gamma\geq10^2$ has been identified with the P-GRB emission ($\Gamma=495$, for GRB 090618, $\Gamma=143$, for GRB 970828 , and $\Gamma=261$ for GRB 101023).

\subsection{The extended afterglow emission}\label{sec:extaft}

After transparency, the residual expanding plasma of leptons and baryons interacts with the CBM and, due to these collisions, starts to slow down giving rise to a multi-wavelength emission: the extended afterglow.
Assuming a fully-radiative condition, the structures observed in the extended afterglow of a GRB are described by two quantities associated with the environment: the CBM density profile $n_{CBM}$, which determines the temporal behavior of the light curve, and the fireshell surface filling factor $\mathcal{R}=A_{eff}/A_{vis}$, in which $A_{eff}$ is the effective emitting area of the fireshell and $A_{vis}$  its total visible area \citep{Ruffini2002,Ruffini2005}. 
This second parameter takes into account the inhomogeneities in the CBM and its filamentary structure \citep{Ruffini2004}.
The emission process of the collision between the baryons and the CBM has been assumed in the comoving frame of the shell as a modified black body spectrum \citep{Patricelli}, given by
\begin{equation}
\label{modBB}
\frac{dN_\gamma}{dVd\epsilon}=\frac{8\pi}{h^3c^3}\left(\frac{\epsilon}{kT}\right)^\alpha\frac{\epsilon^2}{\exp(\epsilon/kT)-1}\ ,
\end{equation}
where $\alpha$ is a phenomenological parameter.
It is appropriate to clarify that this emission is different from the photospheric one due to the $e^+e^-$ plasma annihilation, since it originates from the interactions between the baryons and the CBM in an optically thin regime.

The observed GRB non-thermal spectral shape is then produced by the convolution of a very large number of modified thermal spectra with different temperatures and different Lorentz and Doppler factors.
This convolution is performed over the surfaces of constant arrival time for the photons at the detector \cite[EQuiTemporal Surfaces, EQTS,][]{Bianco2005b,Bianco2005a} encompassing the total observation time. 
The observed hard-to-soft spectral variation comes out naturally from the decrease with time of the comoving temperature and of the bulk Lorentz $\Gamma$ factor. 
This effect is amplified by the curvature effect originated by the EQTS, which produce the observed time lag in the majority of the GRBs.

\begin{figure*}
\centering
\includegraphics[width=0.49\hsize,clip]{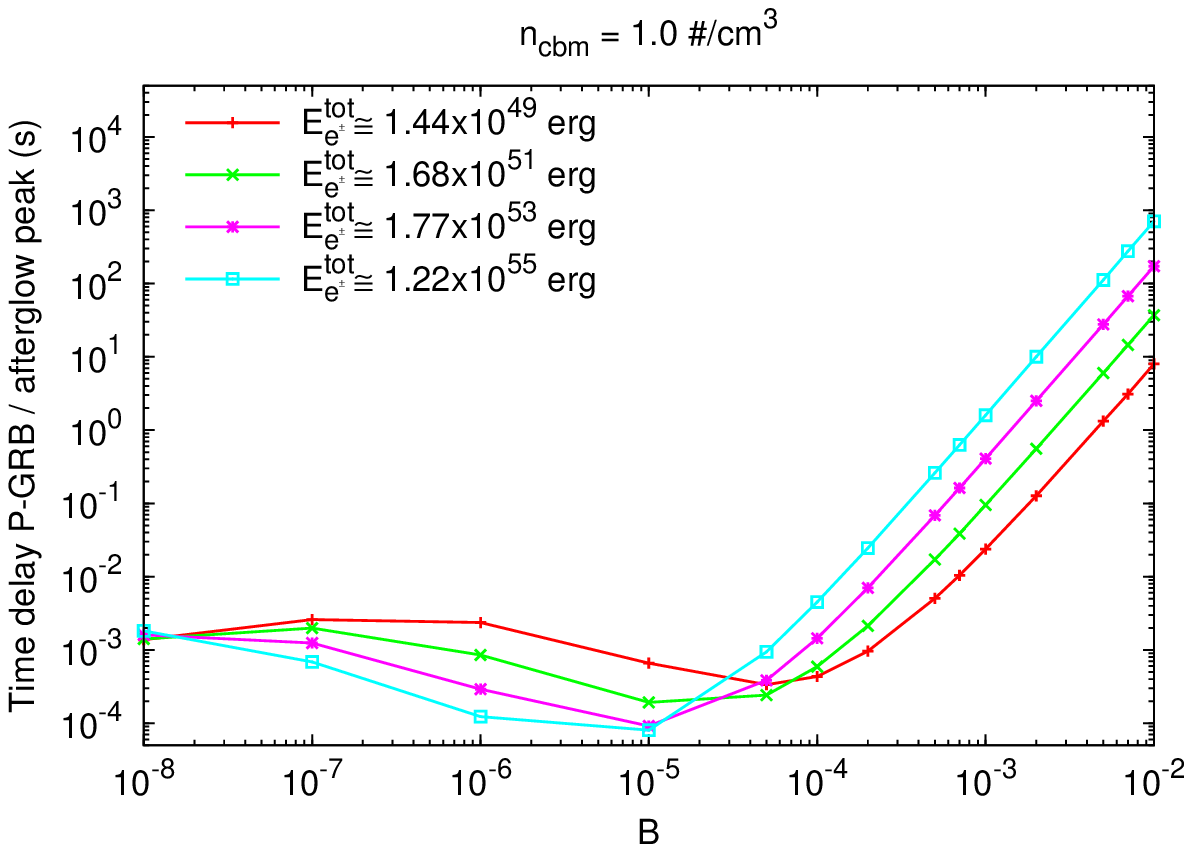}
\includegraphics[width=0.49\hsize,clip]{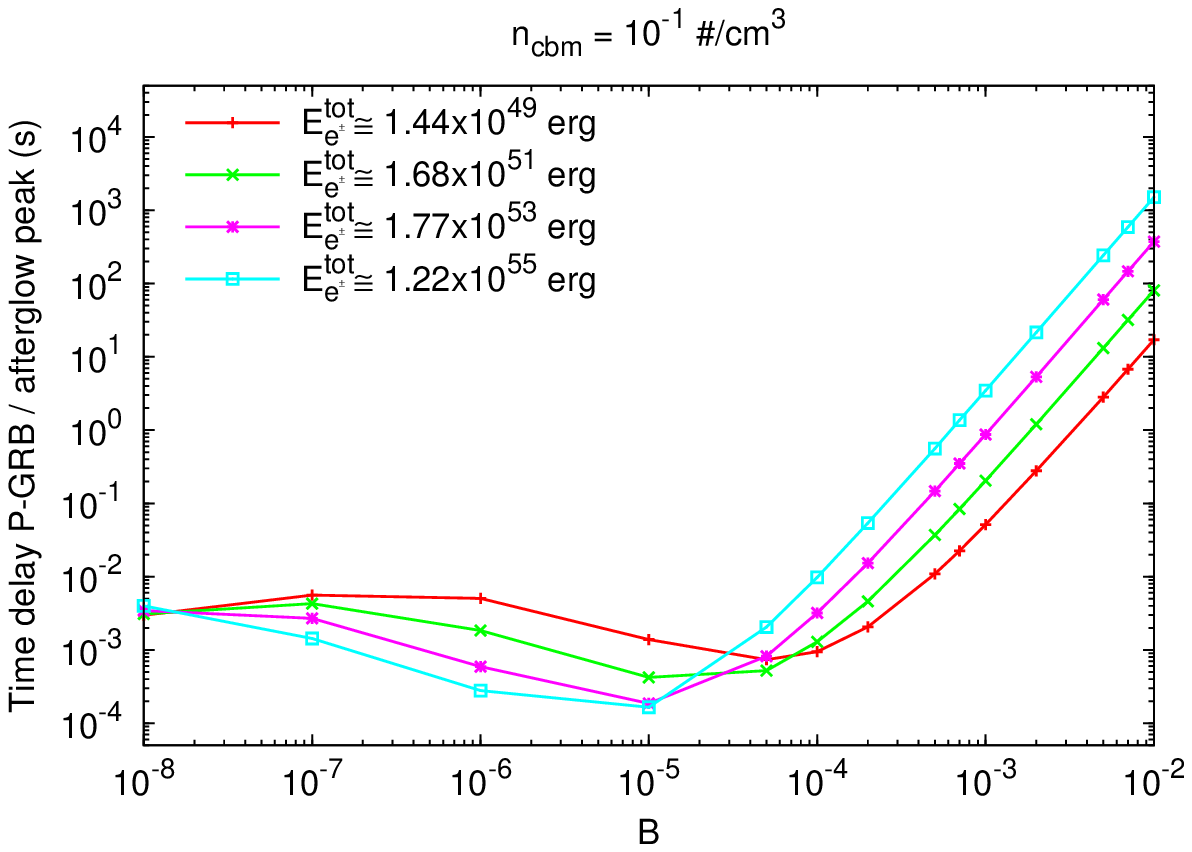}
\includegraphics[width=0.49\hsize,clip]{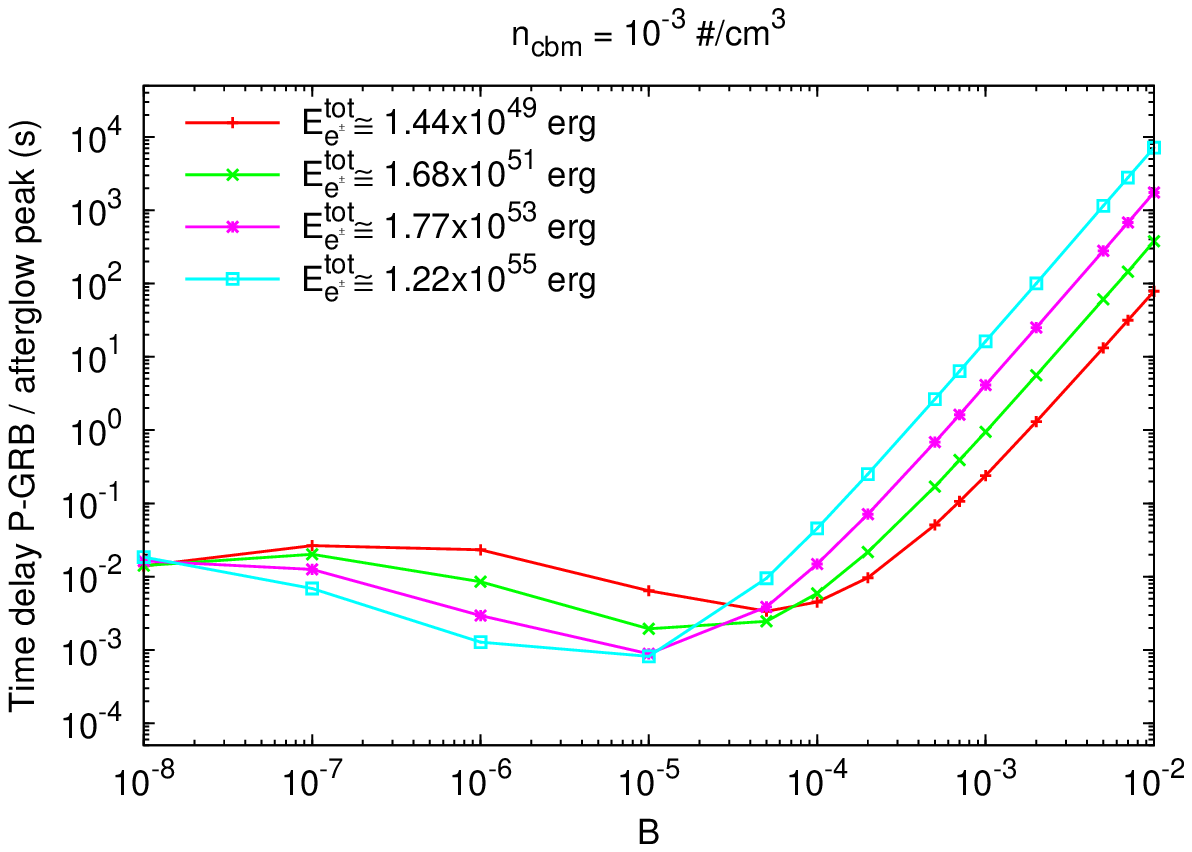}
\includegraphics[width=0.49\hsize,clip]{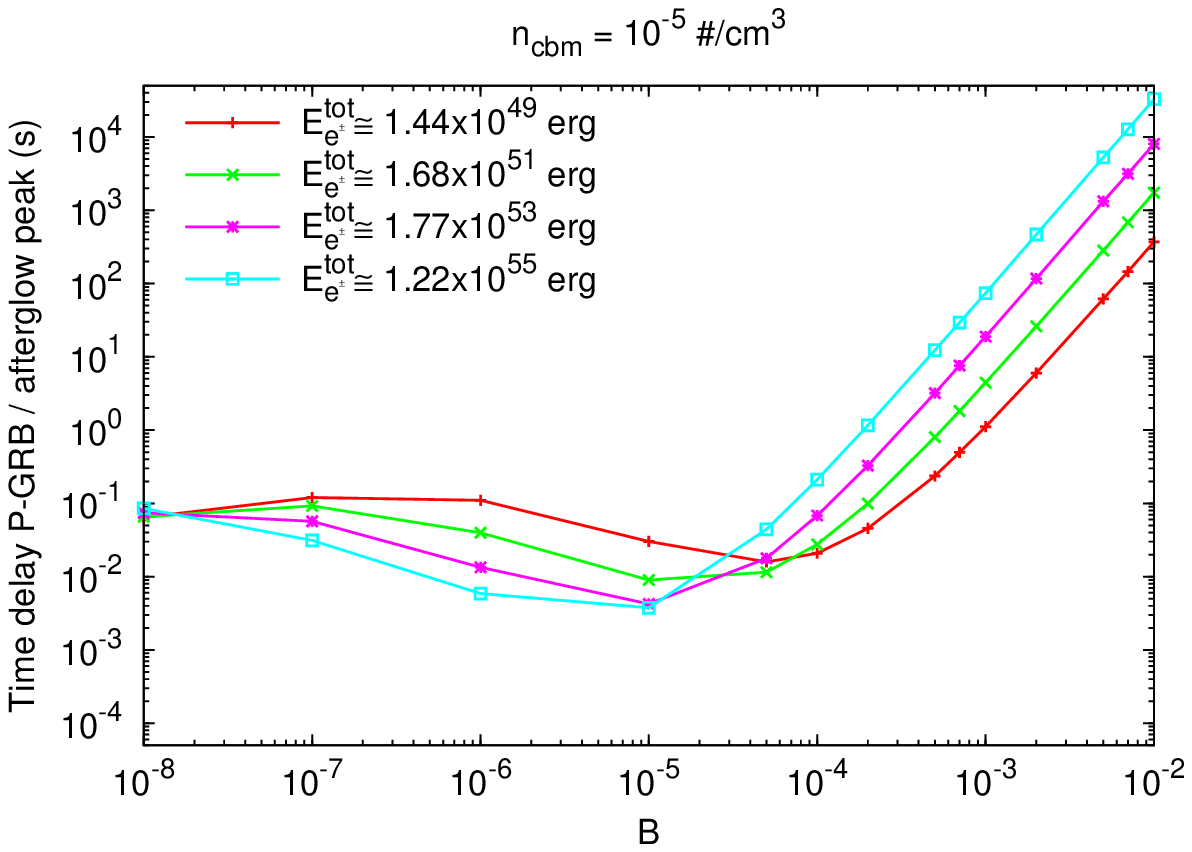}
\caption{Plots of the arrival time separation $\Delta t_a$ between the P-GRB and the peak of the extended afterglow as function of $B$ for four different values of $E_{e^+e^-}^{tot}$, measured in the source cosmological rest frame. This computation has been performed assuming four constant CBM density $n_{CBM} = 1.0,\,1.0\times 10^{-1},\,1.0\times 10^{-3},\,1.0\times 10^{-5}$ particles/cm$^3$.}
\label{fig:2c}
\end{figure*}

\begin{figure}
\centering
\includegraphics[width=\hsize,clip]{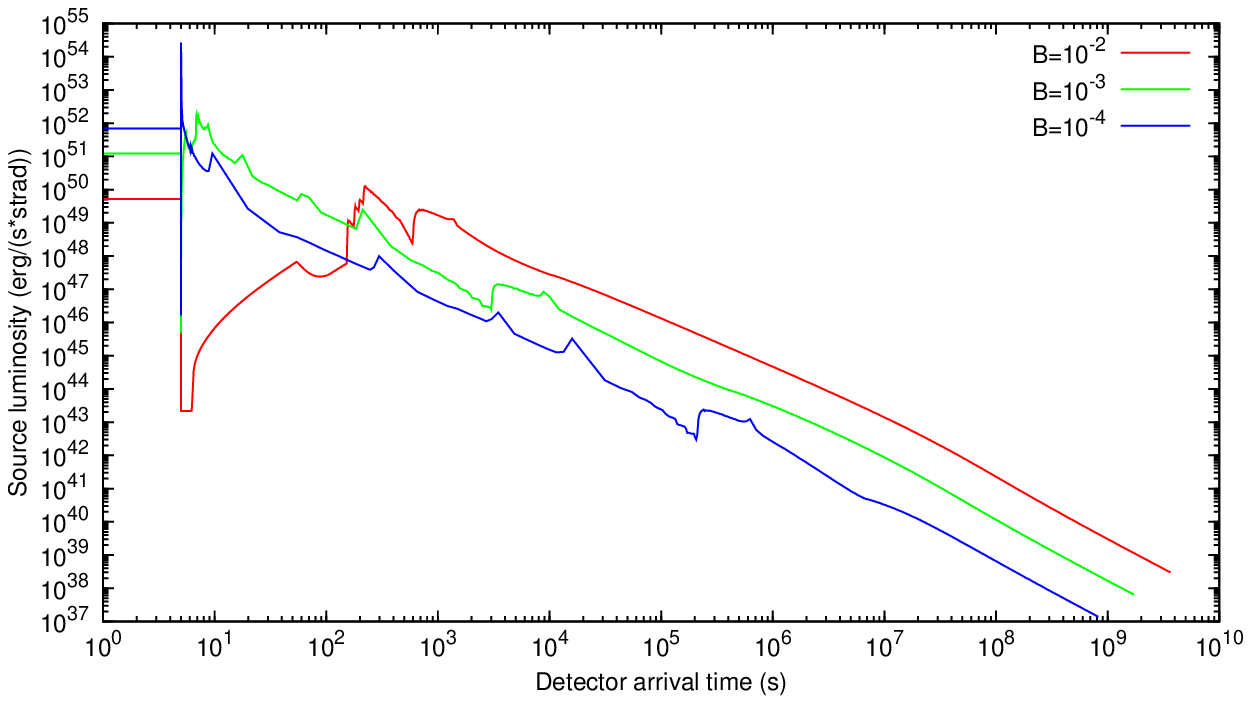} 
\caption{The dependence of the shape of the light curve on $B$. The computations have been performed assuming $E^{tot}_{e^+e^-} = 4.83\times10^{53}$ ergs, $\langle n_{CBM} \rangle = 1.0$ particles/cm$^3$, three different values of the Baryon load $B = 10^{-2},10^{-3},10^{-4}$ and the P-GRBs duration fixed, i.e. $5$ s. For B decreasing, the extended afterglow light curve squeezes itself on the P-GRB and the peak becomes sharper and higher.}
\label{letizia}
\end{figure}

Assuming the spherical symmetry of the system, the isotropic energy emitted in the burst, $E_{iso}$, is equal to the energy of the $e^+e^-$ plasma, $E_{e^+e^-}^{tot}$, and the GRB bolometric light curve is composed of the P-GRB and the extended afterglow. 
Their relative energetics and observed time separation are functions of the energy $E_{e^+e^-}^{tot}$, of the Baryon load $B$, and of the CBM density distribution $n_{CBM}$ (see Fig.~\ref{fig:2c}). 
In particular, for $B$ decreasing, the extended afterglow light curve ``squeezes'' itself on the P-GRB and the P-GRB peak luminosity increases (see Fig.~\ref{letizia}).

To reproduce the shape of the light curve we have to determine for each CBM clump the filling factor $\mathcal{R}$, which determines the effective temperature in the comoving frame and the corresponding peak energy of the spectrum, and of the CBM density $n_{CBM}$, which determines the temporal behavior of the light curve.
It is clear that, since the EQTS encompass emission processes occurring at different comoving times weighted by their Lorentz and Doppler factors, the fit of a single spike is not only a function of the properties of the specific CBM clump but of the entire previous history of the source. 
Due to the non-linearity of the system and to the EQTS, any change in the simulation produces observable effects up to a much later time.
This brings to an extremely complex procedure by trial and error in the data simulation to reach the uniqueness. 

It is appropriate to recall that in the Fireshell model the two phases, the one preceding the $e^+e^-$ transparency and the following one, as well as their corresponding energetics, are directly linked by the Fireshell equations of motion (see Fig.~\ref{fig:4}).
Consequently, their agreement with the data cannot be independently adjusted and optimized.

\subsection{The canonical long GRBs}\label{sec:fireshell:long}

According to this theory, the canonical long GRBs are characterized by a Baryon load varying in the range $3.0\times10^{-4} \lesssim B \leq 10^{-2}$ and they occur in a typical galactic CBM with an average density $\langle n_{CBM} \rangle \approx 1$ particle/cm$^3$.
As a result the extended afterglow is predominant with respect to the P-GRB (see Fig.~\ref{fig:1}).

\subsection{The disguised short GRBs}\label{sec:fireshell:disguised}

After the observations by Swift of GRB 050509B \citep{Gehrels2005}, which was declared in the literature as the first short GRB with an extended emission ever observed, it has become clear that all such sources are actually disguised short GRBs \citep{deBarros2011}. 
It is conceivable and probable that also a large fraction of the declared short duration GRBs in the BATSE catalog, observed before the discovery of the afterglow, are members of this class.
In the case of the disguised short GRBs the Baryon load varies in the same range of the long bursts, while the CBM density is of the order of $10^{-3}$ particles/cm$^3$. 
As a consequence, the extended afterglow results in a ``deflated'' emission that can be exceeded in peak luminosity by the P-GRB \citep{Bernardini2007,Bernardini2008,Caito2009, Caito2010, deBarros2011}. 
Indeed the integrated emission in the extended afterglow is much larger than the one of the P-GRB (see Fig.~\ref{fig:1}), as expected for long GRBs.
With these understandings long and disguised short GRBs are interpreted in terms of long GRBs exploding, respectively, in a typical galactic density or in a galactic halo density.

These sources have given the first evidence of GRBs originating from binary mergers, formed by two neutron stars and/or white dwarfs in all possible combinations, that have spiraled out from their host galaxies into the halos \citep{Bernardini2007,Bernardini2008,Caito2009,Caito2010,deBarros2011}. 
This interpretation has been supported by direct optical observations of GRBs located in the outskirt of the host galaxies \citep{Sahu1997,vanParadijs1997,Bloom2006,Troja2008,Fong2010,Berger2011,Kopaz2012}.

\subsection{The class of genuine short GRBs}\label{sec:fireshell:genuine}

The canonical genuine short GRBs occur in the limit of very low Baryon load, e.g. $B \lesssim 10^{-5}$ with the P-GRB predominant with respect to the extended afterglow.
For such small values of $B$ the afterglow peak emission shrinks over the P-GRB and its flux is lower than the P-GRB one (see Fig.~\ref{letizia}).

The thermalization of photon-pairs plasma is reached in a very short timescale at the beginning of the expansion phase and the thermal equilibrium is implemented during the entire phase of the expansion \citep{2007PhRvL..99l5003A}, therefore the spectrum of these genuine short GRBs is expected to be characterized by a significant thermal-like emission. 
Since the baryon load is small but not zero, in addition to the predominant role of the P-GRB, a non-thermal component originating from the extended afterglow is expected. 

\section{Observations and Data Analysis of GRB 090227B}\label{sec:analysis}

At 18:31:01.41 UT on 27$^{th}$ February 2009, the Fermi GBM detector \citep{GCN8921} triggered and located the short and bright burst, GRB 090227B (trigger 257452263/090227772). The on-ground calculated location, using the GBM trigger data, was (RA,\,Dec)(J2000)=(11$^h$48$^m$36$^s$,\,32$^{\rm{o}}10'12''$), with an uncertainty of 1.77$^{\rm{o}}$ (statistical only).  
The angle from the Fermi LAT boresight was 72$^{\rm{o}}$. 
The burst was also located by IPN \citep{GCN8925} and detected by Konus-Wind \citep{GCN8926}, showing a single pulse with duration $\sim 0.2$ s ($20$ keV -- $10$ MeV).
No X rays and optical observations were reported on the GCN Circular Archive, thus the redshift of the source is unknown. 

To obtain the Fermi GBM light-curves and the spectrum in the energy range $8$ keV -- $40$ MeV, we made use of the \texttt{RMFIT} program.
For the spectral analysis, we have downloaded from the gsfc website \footnote{ftp://legacy.gsfc.nasa.gov/fermi/data/gbm/bursts} the \texttt{TTE} (Time-Tagged Events) files, suitable for short or highly structured events.
We used the light curves corresponding to the NaI-n2 ($8$ -- $900$ keV) and the BGO-b0 ($250$ keV -- $40$ MeV) detectors. 
The $64$ ms binned GBM light curves show one very bright spike with a short duration of $0.384$ s, in the energy range $8$ keV -- $40$ MeV, and a faint tail lasting up to $0.9$ s after the trigtime $T_0$ in the energy range $10$ keV -- $1$ MeV.
After the subtraction of the background, we have proceeded with the time-integrated and time-resolved spectral analyses.

\subsection{Time-integrated spectral analysis}\label{sec:timeint}

\begin{figure*}
\centering 
\hfill
\includegraphics[width=0.42\hsize,clip]{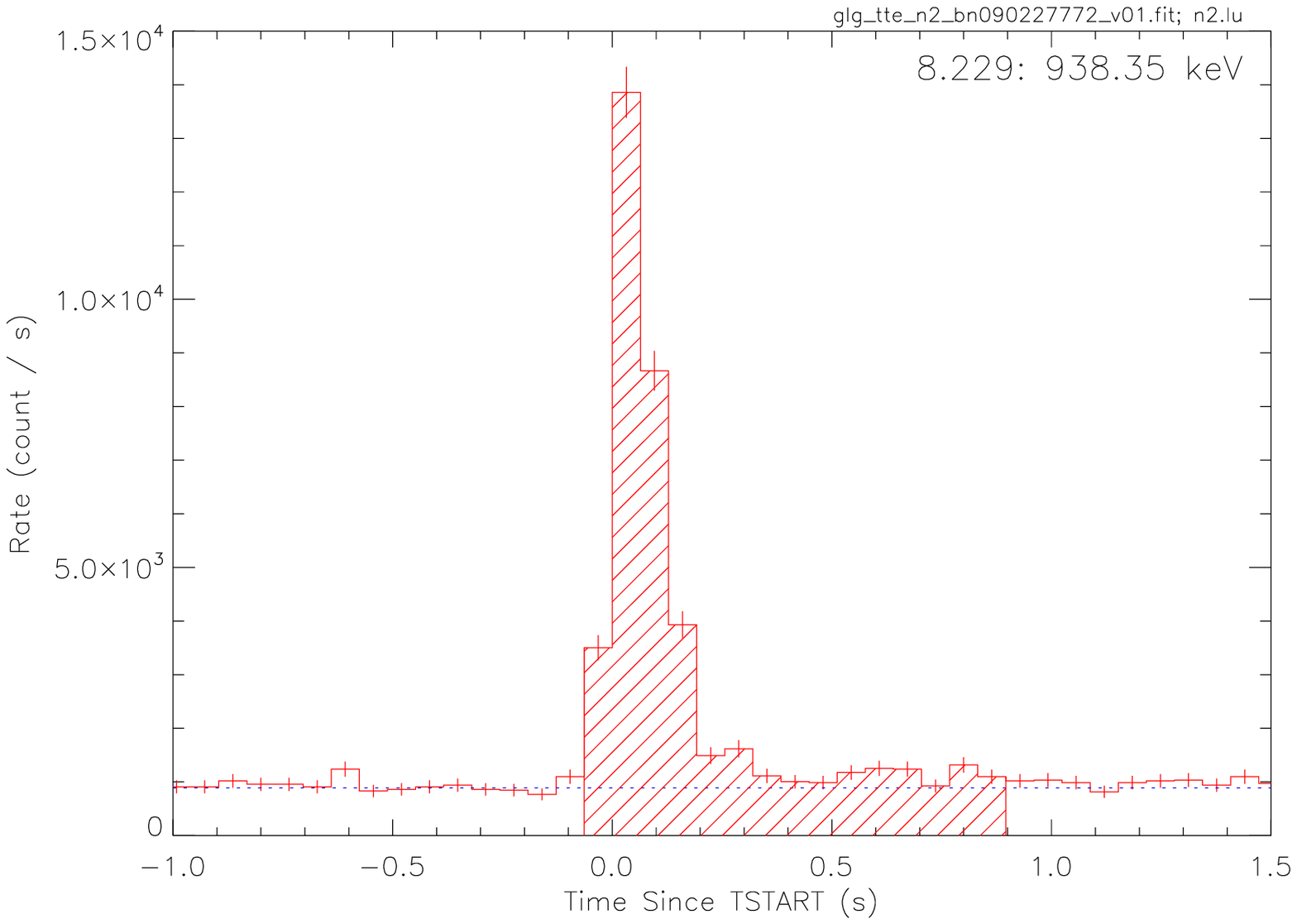}
\hfill
\includegraphics[width=0.42\hsize,clip]{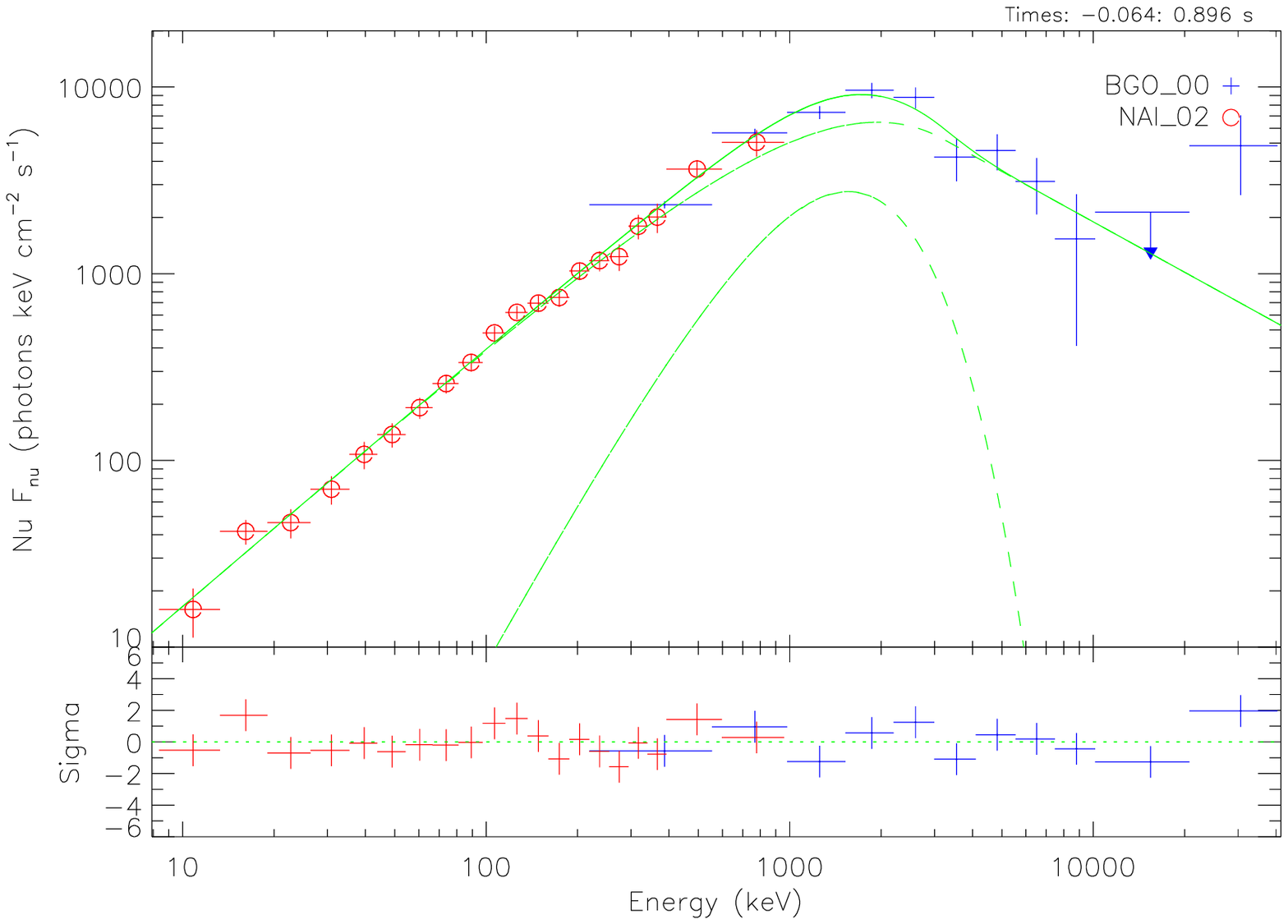} 
\hfill\null\\
\hfill
\includegraphics[width=0.42\hsize,clip]{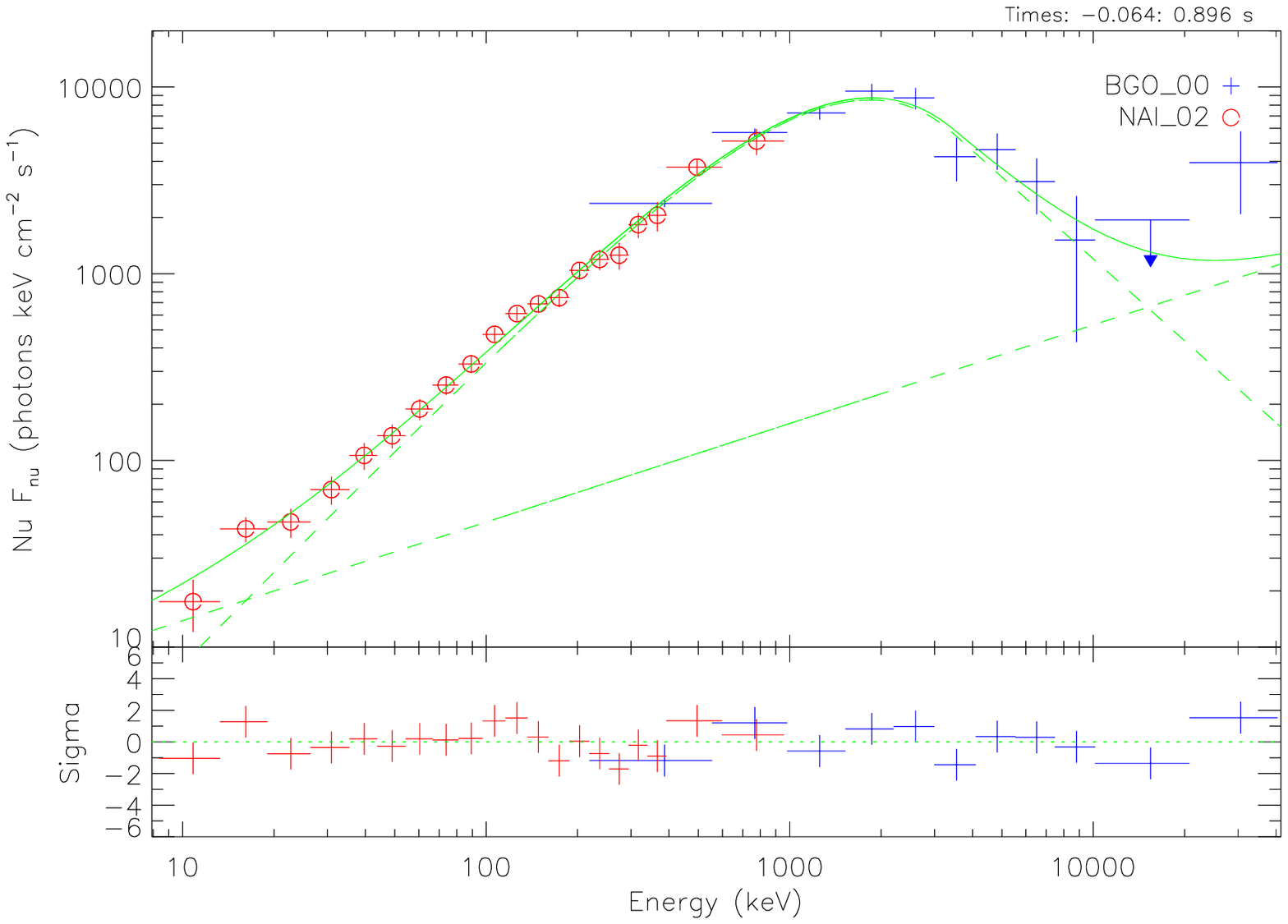}
\hfill
\includegraphics[width=0.42\hsize,clip]{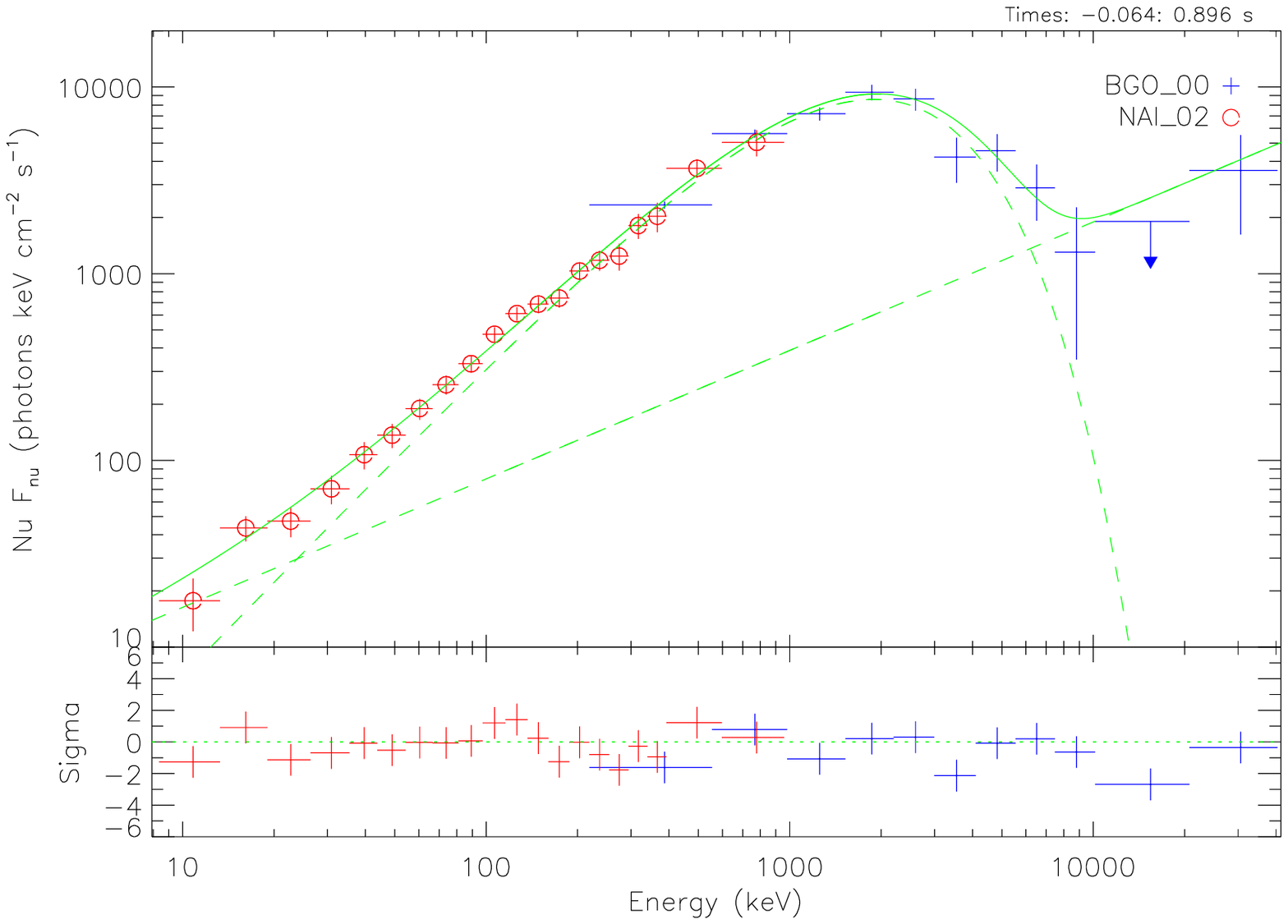} 
\hfill\null\\
\caption{The 64 ms time-binned NaI-n2 light curve (top left panel) and the NaI-n2+BGO-b0 $\nu F_\nu$ spectra (top right BB+Band, bottom left Band+PL, bottom right Compt+PL) of GRB 090227B in the $T_{90}$ time interval.}
\label{fig:2a}
\end{figure*}

We have performed a time-integrated spectral analysis in the time interval from $T_0-0.064$ s to $T_0+0.896$ s, which corresponds to the $T_{90}$ duration of the burst. 
We have fitted the spectrum in this time interval considering the following models: Comptonization (Compt) plus power-law (PL) and Band \citep{Band1993} plus PL, as outlined e.g. in \citet{Guiriec2010}, as well as a combination of Black Body (BB) and Band.
We have evaluated the significance values from the differences in the C-STAT, considered as $\chi^2$ variables for the change in the number of the model parameters.
In Tab.~\ref{sign} we have compared the model with different number of degrees of freedom (DOF). 
Within the $T_{90}$ time interval, the BB+Band model improves the fit with respect to Compt+PL model at a significance level of $5\%$. 
The comparison between Band+PL and Compt+PL models is outside of such a confidence level (about $8\%$). 
The direct comparison between BB+Band and Band+PL models, which have the same number of dof (see Tab.~\ref{sp}), provides almost the same C-STAT values for BB+Band and Band+PL models ($\Delta\textnormal{C-STAT}\approx0.89$).
This means that all the three models are viable. 
The results of the analysis are shown in Tab.~\ref{sp} and Fig.~\ref{fig:2a}. 
For BB+Band model, the ratio between the fluxes of the thermal component and the non-thermal one (NT) is $F_{BB}/F_{NT} \approx 0.22$.
The BB component is important in the determination of the peak of the $\nu F_\nu$ spectrum and has an observed temperature $kT = (397\pm70)$ keV.

\begin{figure*}
\centering 
\hfill
\includegraphics[width=0.42\hsize,clip]{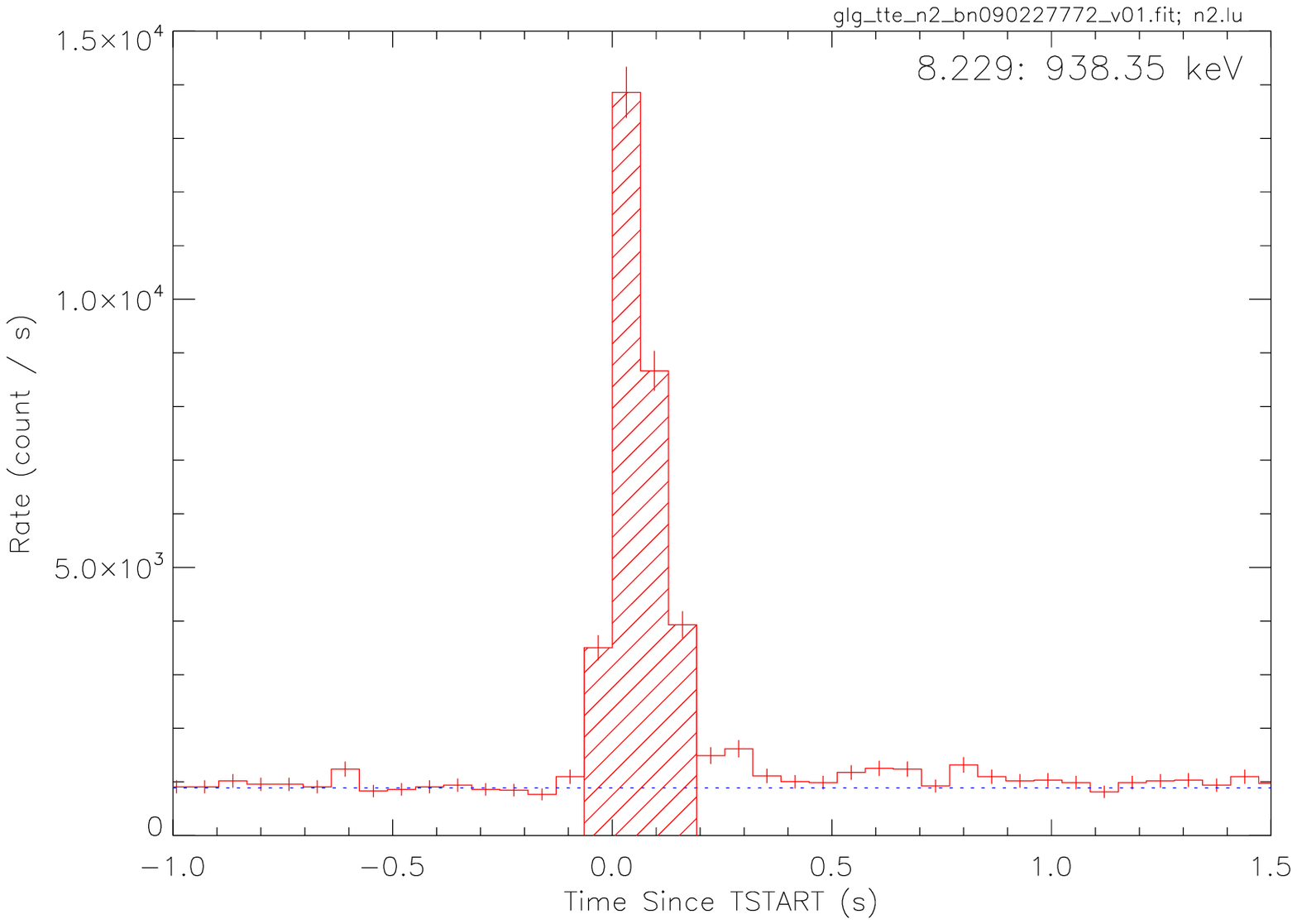}
\hfill
\includegraphics[width=0.42\hsize,clip]{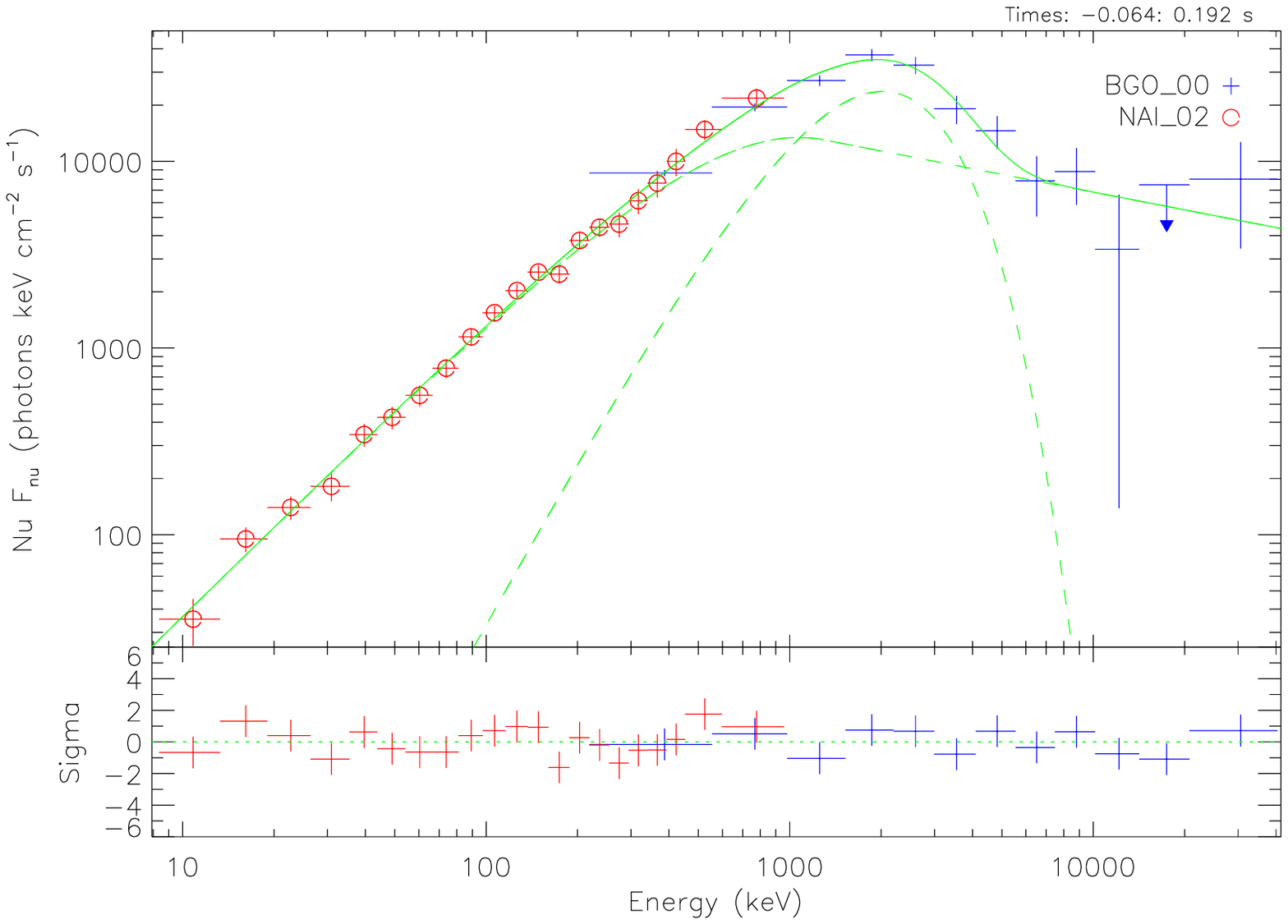} 
\hfill\null\\
\hfill
\includegraphics[width=0.42\hsize,clip]{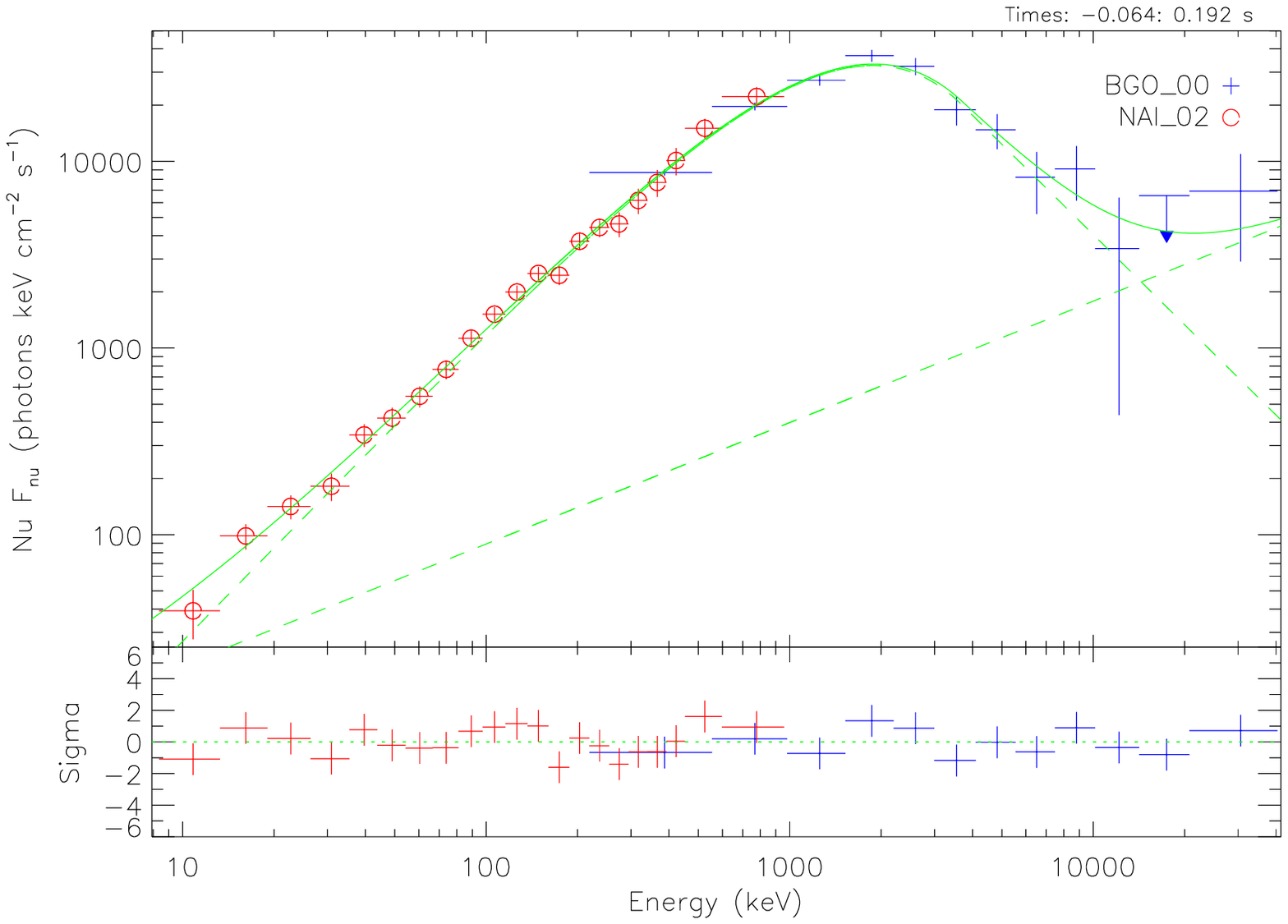}
\hfill
\includegraphics[width=0.42\hsize,clip]{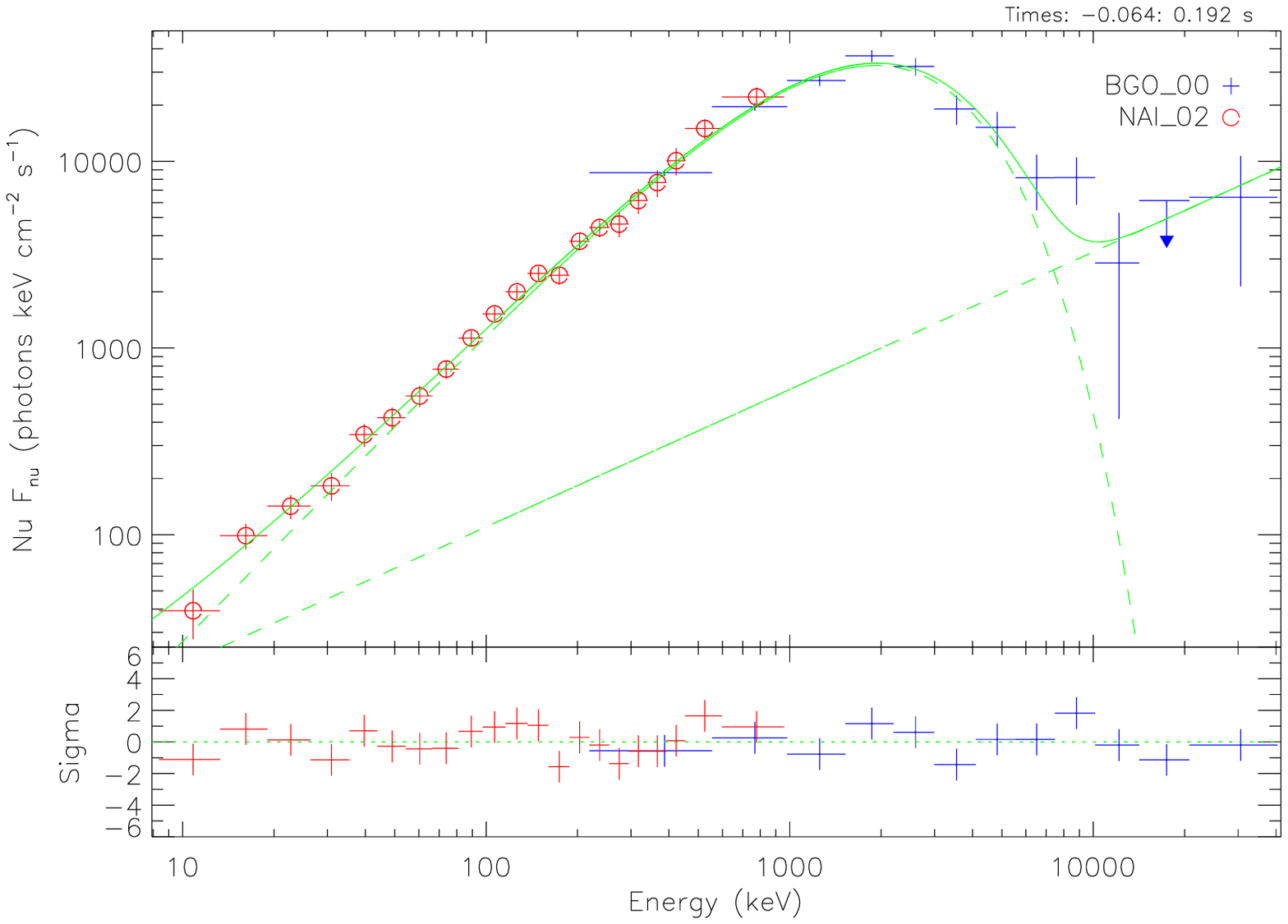}
\hfill\null\\
\caption{The same considerations as in Fig.~\ref{fig:2a}, in the $T_{spike}$ time interval.}
\label{fig:2aa}
\end{figure*}

We have then focused our attention on the spike component, namely the time interval from $T_0-0.064$ s to $T_0+0.192$, which we indicate in the following as the $T_{spike}$.
We have repeated the time-integrated analysis considering the same spectral models of the previous interval (see Tab.~\ref{sp} and Fig.~\ref{fig:2aa}).
As reported in Tab.~\ref{sign}, within the $T_{spike}$ time interval, both BB+Band and Band+PL models marginally improve the fits of the data with respect to Compt+PL model within a confidence level of $5\%$. 
Again, the C-STAT values of BB+Band and Band+PL models are almost the same ($\Delta\textnormal{C-STAT}\approx0.15$) and they are statically equivalent in the $T_{spike}$.
For the BB+Band model, the observed temperature of the thermal component is $kT = (515\pm28)$ keV and the flux ratio between the BB component and the NT one increases up to $F_{BB}/F_{NT} \approx 0.69$.

\begin{figure*}
\centering 
\hfill
\includegraphics[width=0.42\hsize,clip]{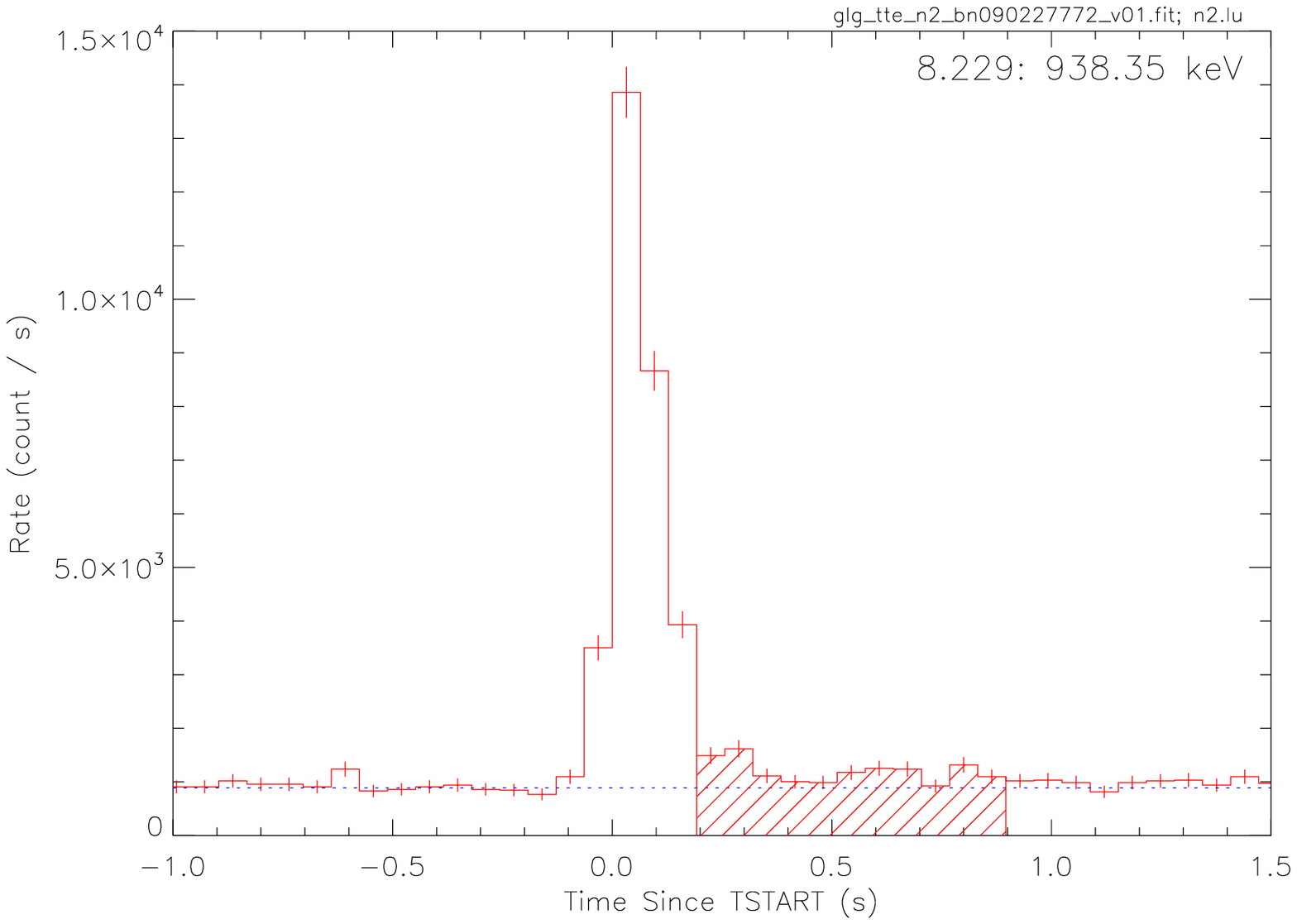}
\hfill
\includegraphics[width=0.42\hsize,clip]{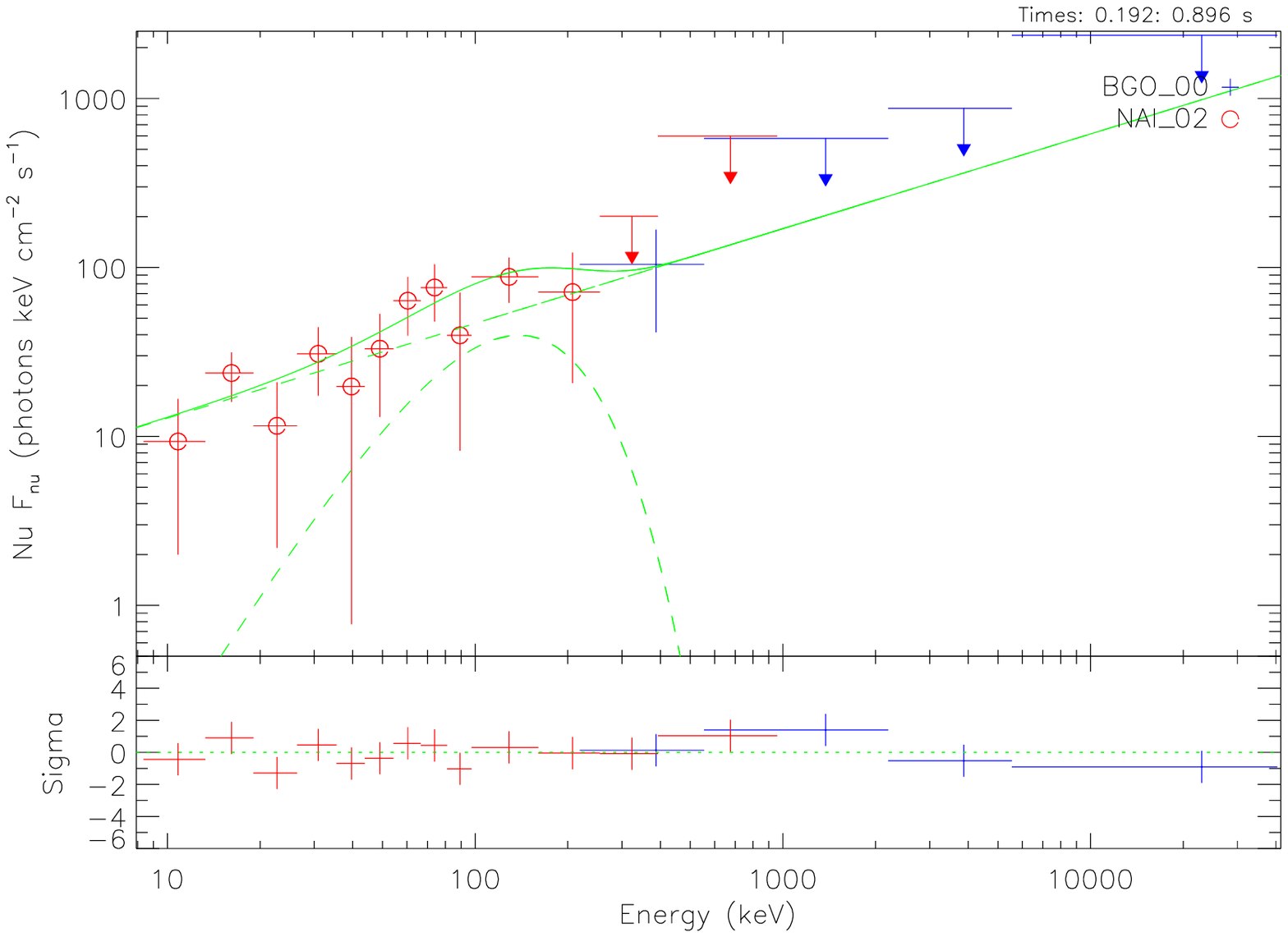} 
\hfill\null\\
\hfill
\includegraphics[width=0.42\hsize,clip]{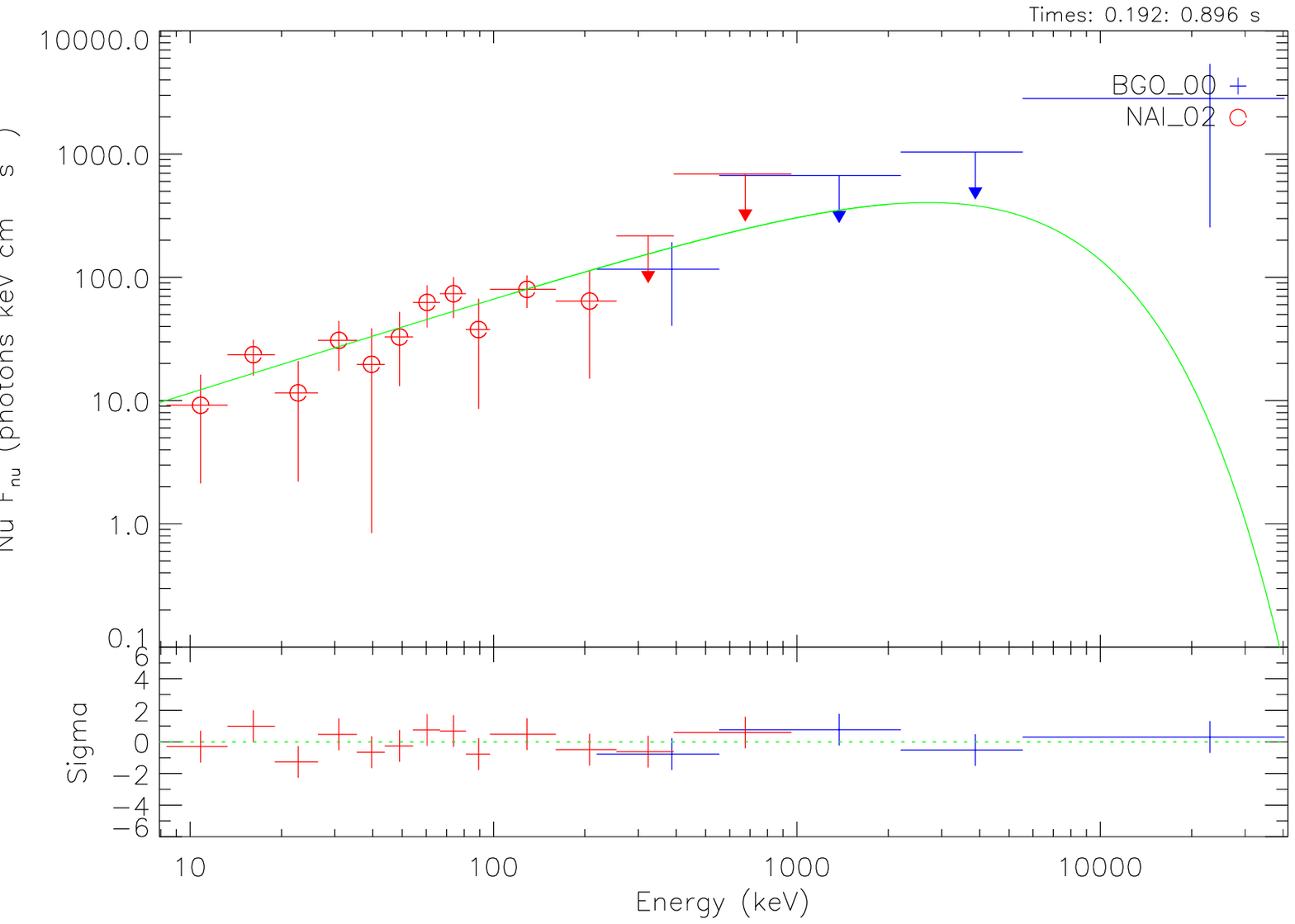}
\hfill
\includegraphics[width=0.42\hsize,clip]{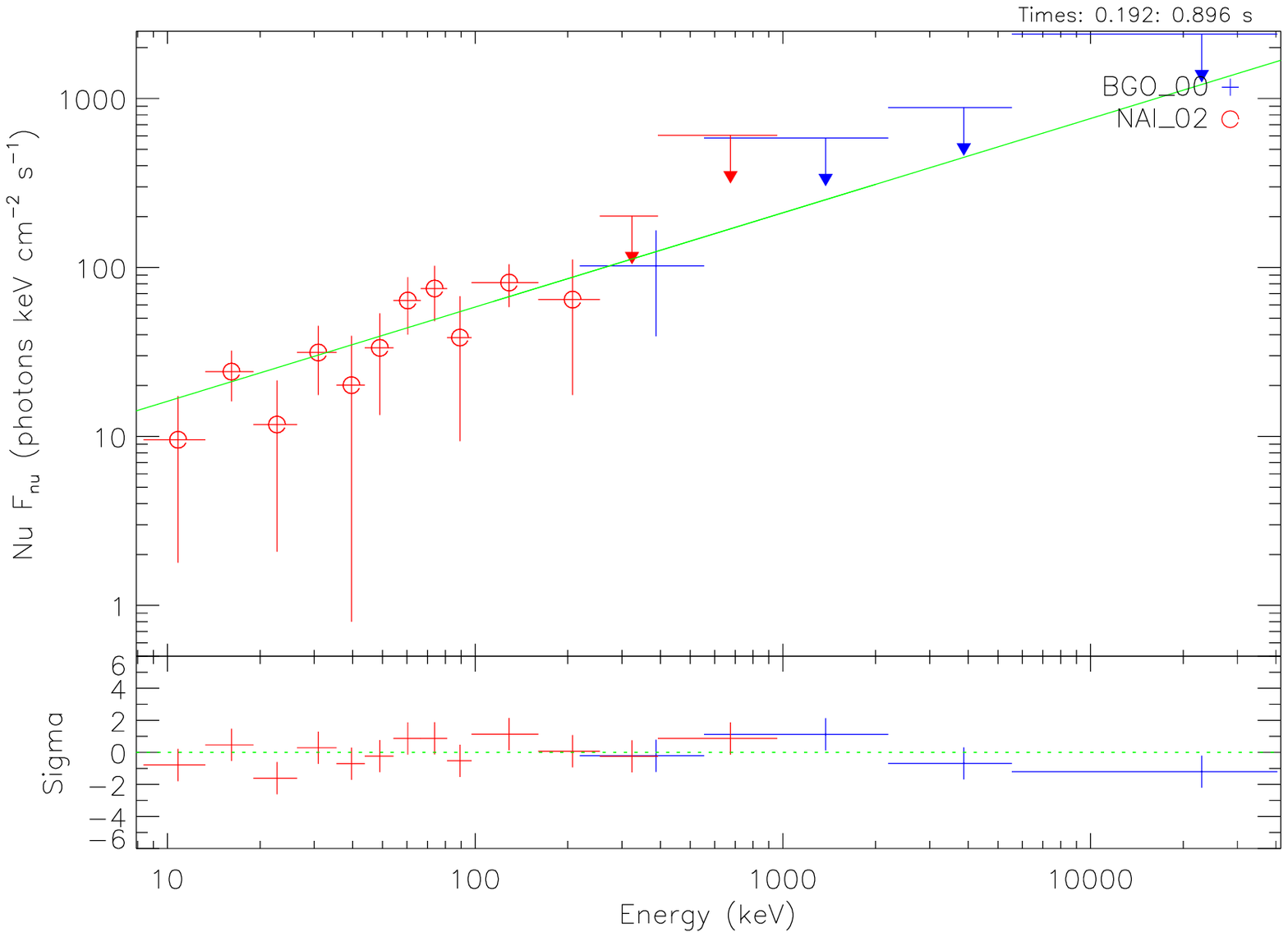} 
\hfill\null\\
\caption{The 64 ms time-binned NaI-n2 light curve (top left panel) and the NaI-n2+BGO-b0 $\nu F_\nu$ spectra (top right BB+PL, bottom left Compt, bottom right PL) of GRB 090227B in the $T_{tail}$ time interval.}
\label{fig:2aaa}
\end{figure*}

We have performed a further analysis in the time interval from $T_0+0.192$ s to $T_0+0.896$ s, which we indicate as $T_{tail}$, by considering BB+PL, Compt and PL models (see Fig.~\ref{fig:2aaa} and Tab.~\ref{sp}). 
The comparison in Tab.~\ref{sign} shows that the best fit is the Compt model.
The BB+PL model is the less preferred.
From the data analysis in the $T_{tail}$ time interval, we can conclude that a thermal component is ruled out.

In view of the above, we have focused our attention on the fit of the data of the BB+Band model within the Fireshell scenario, been equally probable from a mere statistical point of view with the other two choices, namely Band+PL and Compt+PL.
According to the Fireshell scenario (see Sec.~\ref{sec:extaft}), the emission within the $T_{spike}$ time interval is related to the P-GRB and is expected to be thermal.
In addition the transition between the transparency emission of the P-GRB and the extended afterglow is not sharp.
The time separation between the P-GRB and the peak of the extended afterglow depends on the energy of the $e^+e^-$ plasma $E_{e^+e^-}^{tot}$, the Baryon load $B$ and the CBM density $n_{CBM}$ (see Fig.~\ref{letizia}).
As shown in Figs.~\ref{fig:2c} and \ref{letizia}, for decreasing values of $B$ an early onset of the extended afterglow in the P-GRB spectrum occurs and thus, a NT component in the $T_{spike}$ is expected.
As a further check, the theory of the Fireshell model indeed predicts in the early part of the prompt emission of GRBs a thermal component due to the transparency of the $e^+e^-$ plasma (see Sec.~\ref{sec:fireshell}), while in the extended afterglow no thermal component is expected (see Sec.~\ref{sec:extaft}), as observed in the $T_{tail}$ time interval.

Our theoretical interpretation is consistent with the observational data and the statistical analysis. 
From an astrophysical point of view the BB+Band model is preferred over the other two models, statistically equivalent in view of the above theoretical considerations.

\begin{table*}
\tiny
\centering
\begin{tabular}{cccccccccc}
\hline\hline
\textbf{Int.}  &  \textbf{Model}  &  \textnormal{\textbf{$kT$ [keV]}}  &  \textnormal{\textbf{$E_{p}$ [keV]}}  &  \textbf{$\alpha$}  &  \textbf{$\beta$}  &  \textbf{$\gamma$}  &  \textbf{$F_{tot}$ [erg/cm$^2$s]}  &  \textnormal{\textbf{$F_{BB}/F_{tot}$}}  &  \textbf{C-STAT/dof} \\
\hline
${}$       &  \textnormal{\textbf{BB+Band}}   &  $397 \pm 70$  &  $1942 \pm 249$  &  $-0.60 \pm 0.05$  &  $-2.90 \pm 0.31$  &  {}                  &  $(3.35 \pm 0.12)\times 10^{-5}$  &  $0.22$  &  $286.84/240$  \\
$T_{90}$   &  \textnormal{\textbf{Band+PL}}     &  {}                &  $1835 \pm 84$   &  $-0.35 \pm 0.05$  &  $-3.46 \pm 0.46$  &  $-1.47 \pm 0.13$    &  $(3.39 \pm 0.13)\times 10^{-5}$  &  ${}$    &  $287.73/240$  \\
${}$  &  \textnormal{\textbf{Compt+PL}}  &  ${}$              &  $1877 \pm 72$   &  $-0.36 \pm 0.05$  &  ${}$                   &  $-1.36 \pm 0.05$  &  $(3.44 \pm 0.13)\times 10^{-5}$  &  ${}$    &  $290.71/241$  \\
\hline
${}$  &  \textnormal{\textbf{BB+Band}}   &  $515 \pm 28$  &  $1072 \pm 210$  &  $-0.40 \pm 0.05$         &  $-2.32 \pm 0.17$  &  {}                  &  $(1.26 \pm 0.04)\times 10^{-4}$  &  $0.69$  &  $266.17/240$  \\
$T_{spike}$   &  \textnormal{\textbf{Band+PL}}     &  {}                &  $1879 \pm 67$   &  $-0.33 \pm 0.05$  &  $-3.61 \pm 0.38$  &  $-1.35 \pm 0.10$  &  $(1.25 \pm 0.04)\times 10^{-4}$  &  ${}$    &  $266.32/240$  \\
${}$  &  \textnormal{\textbf{Compt+PL}}  &  ${}$              &  $1912 \pm 58$   &  $-0.33 \pm 0.05$  &  ${}$              &  $-1.26 \pm 0.07$  &  $(1.26 \pm 0.04)\times 10^{-4}$  &  ${}$    &  $270.19/241$  \\
\hline
${}$  &  \textnormal{\textbf{BB+PL}}   &  $36 \pm 13$  &                                   &                    &          &  $-1.44 \pm 0.07$           &  $(3.9 \pm 1.2)\times 10^{-6}$  &  $unc.$  &  $293.85/242$  \\
$T_{tail}$    &  \textnormal{\textbf{Compt}}     &  {}                &  $2703 \pm 1760$   &  $-1.23 \pm 0.09$  &         &                             &  $(2.03 \pm 0.79)\times 10^{-6}$  &  ${}$    &  $291.19/243$  \\
${}$  &  \textnormal{\textbf{PL}}  &  ${}$              &                                  &                    &  ${}$    &  $-1.44 \pm 0.05$  &  $(4.7 \pm 1.1)\times 10^{-6}$  &  ${}$    &  $296.07/244$  \\
\hline
\end{tabular}
\caption{The time-integrated spectral analyses performed using BB+Band, Band+PL and Compt+PL models in the $T_{90}$ and $T_{spike}$ time intervals, and BB+PL, Compt and PL in the $T_{tail}$ time interval, in the energy range $8$ keV -- $40$ MeV.}
\label{sp}
\end{table*}
\begin{table}
\tiny
\centering
\begin{tabular}{cccc}
\hline\hline
\textbf{Int.}  &  \textbf{Models}                  &  \textbf{$\Delta$}\textbf{C-STAT}  &  \textbf{Significance}  \\
\hline
$T_{90}$       &  \textbf{BB+Band over Compt+PL}   &  $3.87$                            &  $0.049$                \\
${}$           &  \textbf{Band+PL over Compt+PL}   &  $2.98$                            &  $0.084$                \\
\hline
$T_{spike}$    &  \textbf{BB+Band over Compt+PL}   &  $4.02$                            &  $0.045$                \\
${}$           &  \textbf{Band+PL over Compt+PL}   &  $3.87$                            &  $0.049$                \\
\hline
${}$           &  \textbf{BB+PL   over       PL}   &  $2.22$                            &  $0.33$                 \\
$T_{tail}$     &  \textbf{BB+PL   over    Compt}   &  $2.66$                            &  $0.10$                 \\
${}$           &  \textbf{Compt   over       PL}   &  $4.88$                            &  $0.027$                \\
\hline
\end{tabular}
\caption{The C-STAT improvement with the addition of extra parameters in the $T_{90}$, $T_{spike}$ and $T_{tail}$ time intervals (see Tab.~\ref{sp}).}
\label{sign}
\end{table} 

\subsection{Time-resolved spectral analysis}\label{sec:timeres}

We have performed a time-resolved spectral analysis on shorter selected time intervals of $32$ ms in order to correctly identify the P-GRB, namely finding out in which time interval the thermal component exceeds or at least has a comparable flux with respect to the NT one due to the onset of the extended afterglow. 
In this way we can single out the contribution of the NT component in the spectrum of the P-GRB.

\begin{figure}
\centering
\includegraphics[width=0.9\hsize,clip]{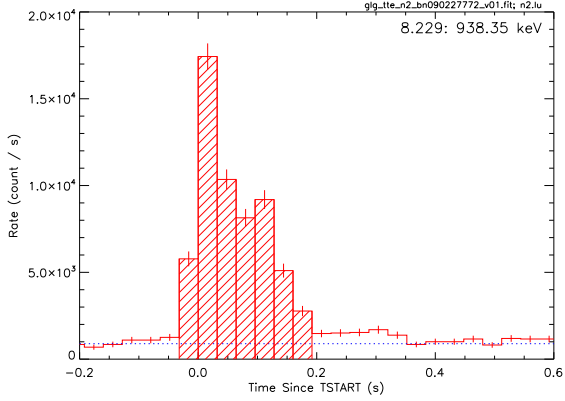}
\caption{The $32$ ms time-binned NaI-n2 light curve of GRB 090227B in the time interval from $T_0-0.032$ s to $T_0+0.192$ s; each time bin corresponds to the time-resolved interval considered in the Sec.~\ref{sec:timeres}.}
\label{fig:timeres}
\end{figure}

A time-resolved spectral analysis has been performed by \citet{Guiriec2010} by selecting time intervals from $2$ ms to $94$ ms.
In view of the low statistical content in some small time bins, the authors fitted the data by using simple Band functions.
We have performed a time-resolved analysis on time intervals of $32$ ms (see Fig.~\ref{fig:timeres}) in order to optimize the statistical content in each time bin and to test the presence of BB plus an extra NT component.
The results are summarized in Tab.~\ref{parint}, where we have compared the BB+NT with the single Band function.

\begin{table*}
\tiny
\centering
\begin{tabular}{cccccccccc}
\hline\hline
\textbf{Interval}         &  \textbf{Models}   & $kT$            &  $E_{p}$        & \textbf{$\alpha$} & \textbf{$\beta$} & $F_{tot}\times10^{-5}$  & \textbf{$\chi^2/DOF$}  & \textbf{$F_{BB}/F_{NT}$}  &  \textbf{BB+Band}   \\
\textbf{[s]}              &                    & \textbf{[keV]}  & \textbf{[keV]}  &                   &                   & \textbf{[erg/cm$^2$s]}  &                        &                        &  \textbf{over Band}  \\
\hline
                          &  \textbf{BB+PL}    &  $274\pm17$     &                 &  $-1.75\pm0.29$   &                   &  $7.03\pm0.76$          &  $196.85/241=0.82$     &  $11.2\pm3.4$           &                          \\
$-0.032\rightarrow0.000$  &  \textbf{BB+Band}  &  $280\pm66$     &  $1703\pm407$   &  $-0.50\pm0.25$   &  unc             &  $8.22\pm0.99$           &  $180.23/239=0.75$     &  $0.50\pm0.26$           &  $0.051$                 \\
                          &  \textbf{Band}     &                 &  $1493\pm155$   &  $-0.21\pm0.11$   &  unc             &  $8.13\pm0.88$           &  $186.17/241=0.77$     &                        &                          \\
\hline
                          &  \textbf{BB+PL}    &  $377\pm12$     &                 &  $-1.20\pm0.03$   &                  &  $62.2\pm3.6$            &  $308.97/241=1.28$     &  $1.04\pm0.11$           &                          \\
$0.000\rightarrow0.032$   &  \textbf{BB+Band}  &  $571\pm44$     &  $858\pm214$    &  $0.15\pm0.17$    &  $-2.45\pm0.26$   &  $46.2\pm2.3$           &  $222.54/239=0.93$     &  $1.41\pm0.38$           &  $0.041$                 \\
                          &  \textbf{Band}     &                 &  $2140\pm102$   &  $-0.10\pm0.06$   &  $-5.3\pm1.9$     &  $47.2\pm2.1$           &  $228.95/241=0.95$     &                        &                          \\
\hline
                          &  \textbf{BB+PL}    &  $437\pm20$     &                 &  $-1.21\pm0.03$   &                   &  $43.4\pm3.1$           &  $247.41/241=1.03$     &  $1.00\pm0.24$           &                          \\ 
$0.032\rightarrow0.064$   &  \textbf{BB+Band}  &  $572\pm65$     &  $1713\pm1045$  &  $-0.42\pm0.14$   &  $-1.77\pm0.26$   &  $35.0\pm2.6$           &  $222.18/239=0.93$     &  $0.55\pm0.35$           &  $0.081$                 \\ 
                          &  \textbf{Band}     &                 &  $2439\pm257$   &  $-0.29\pm0.07$   &  $-2.64\pm0.22$   &  $36.4\pm2.6$           &  $227.21/241=0.94$     &                        &                          \\
\hline
                          &  \textbf{BB+PL}    &  $329\pm21$     &                 &  $-1.41\pm0.04$   &                   &  $17.8\pm1.9$           &  $241.91/241=1.00$     &  $0.92\pm0.27$           &                          \\
$0.064\rightarrow0.096$   &  \textbf{BB+Band}  &  $373\pm34$     &  $435\pm297$    &  $-0.48\pm0.09$   &  $-1.70\pm0.14$   &  $17.5\pm1.9$           &  $221.50/239=0.93$     &  $0.85\pm0.28$           &  $0.020$                 \\
                          &  \textbf{Band}     &                 &  $1586\pm281$   &  $-0.48\pm0.29$   &  $-2.23\pm0.19$   &  $17.5\pm2.0$           &  $229.31/241=0.95$     &                        &                          \\
\hline
                          &  \textbf{BB+PL}    &  $124.9\pm8.4$  &                 &  $-1.27\pm0.04$   &                   &  $18.9\pm0.23$          &  $258.17/241=1.07$     &  $0.21\pm0.08$           &                          \\
$0.096\rightarrow0.128$   &  \textbf{BB+Band}  &  $144\pm84$     &  $454\pm162$    &  $0.11\pm0.30$    &  $-1.80\pm0.17$   &  $16.1\pm2.1$           &  $226.61/239=0.95$     &  unc                   &  $0.061$                 \\
                          &  \textbf{Band}     &                 &  $622\pm112$    &  $-0.11\pm0.17$   &  $-1.99\pm0.11$   &  $13.8\pm1.8$           &  $232.21/241=0.96$     &                        &                          \\
\hline
                          &  \textbf{BB+PL}    &  $35.5\pm4.8$   &                 &  $-1.52\pm0.08$   &                  &  $2.87\pm0.95$          &  $202.44/241=0.84$     &  $0.13\pm0.06$           &                          \\
$0.128\rightarrow0.160$   &  \textbf{BB+Band}  &  $39.6\pm6.8$   &  unc            &  $-1.2\pm1.4$     &  $-1.54\pm0.26$   &  $2.8\pm1.2$            &  $198.00/239=0.83$     &  $0.14\pm0.08$         &  $0.067$                 \\
                          &  \textbf{Band}     &                 &  $193\pm124$    &  $-0.75\pm0.40$   &  $-1.84\pm0.18$   &  $2.55\pm0.91$          &  $203.40/241=0.84$     &                        &                          \\
\hline
                          &  \textbf{BB+PL}    &  $30.2\pm7.7$   &                 &  $-1.19\pm0.10$   &                   &  $5.7\pm1.4$            &  $237.82/241=0.99$     &  $0.020\pm0.019$       &                          \\
$0.160\rightarrow0.192$   &  \textbf{BB+Band}  &  $22\pm10$      &  unc            &  unc              & $-1.25\pm0.08$    & unc                     &  $203.37/239=0.85$     &  unc                   &  $0.0045$                \\
                          &  \textbf{Band}     &                 &  unc            &  $-0.7\pm1.2$     & $-1.40\pm0.08$    &  $6.0\pm1.4$            &  $214.19/241=0.89$     &                        &                          \\
\hline
\end{tabular}
\caption{Time-resolved analysis of GRB 090227B performed using BB+NT (NT=Band,PL) and a single Band model. In the first column we have indicated the time bin; in the following five columns we have indicated the spectral models and their parameters. In the next three columns we have listed, respectively, the total flux, the $\chi^2$, and the ratio between the thermal (where considered) and the non-thermal fluxes. The last column reports the significance in the addition of the the BB with respect the sole Band model.}
\label{parint}
\end{table*}

In the analysis we have preferred the $\chi^2$ statistic because of the high photon fluxes in the first five time intervals, $\gtrsim 100$ photons/(cm$^2$s).

Within the first time-resolved interval the BB+PL model has a thermal flux $(11.2\pm3.4)$ times bigger than the PL flux; the fit with BB+Band provides $F_{BB}=(0.50\pm0.26)F_{NT}$, where the NT component is in this case the Band model.
In the second and fourth intervals, the BB+Band model provides an improvement at a significance level of $5\%$ in the fitting procedure with respect to the simple Band model (see Tab.~\ref{parint}, last column). 
In the third time interval as well as in the remaining time intervals up to $T_0+0.192$ s the Band spectral models provide better fits with respect to the BB+NT ones. 

This is exactly what we expect from our theoretical understanding: from $T_0-0.032$ s to $T_0+0.096$ s we have found the edge of the P-GRB emission, in which the thermal components have fluxes higher or comparable to the NT ones.
The third interval corresponds to the peak emission of the extended afterglow (see Fig.~\ref{fig:2d1}).
The contribution of the extended afterglow in the remaining time intervals increases, while the thermal flux noticeably decreases (see Tab.~\ref{parint}).

\begin{figure*}
\centering
\hfill
\includegraphics[width=0.42\hsize,clip]{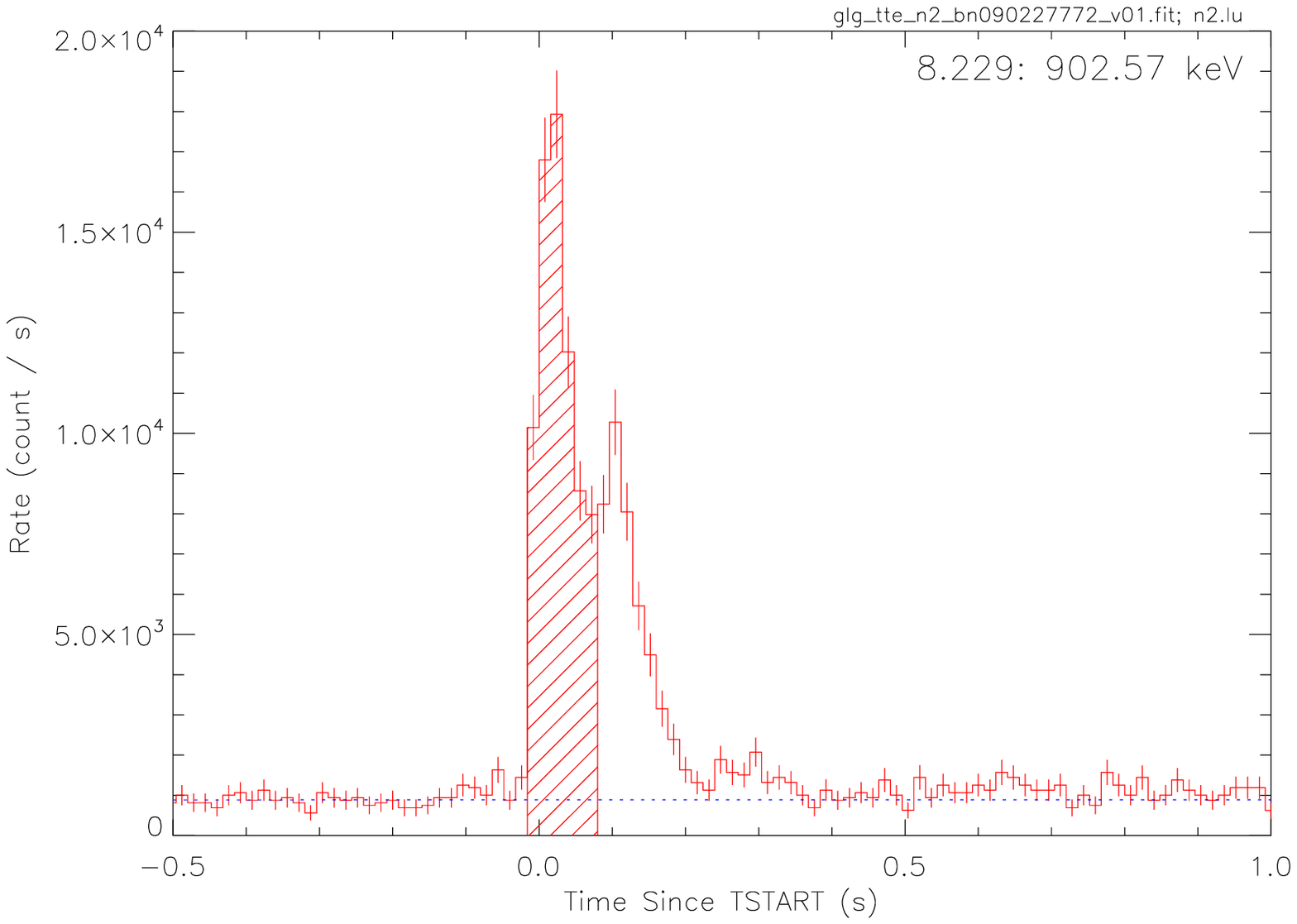}
\hfill
\includegraphics[width=0.42\hsize,clip]{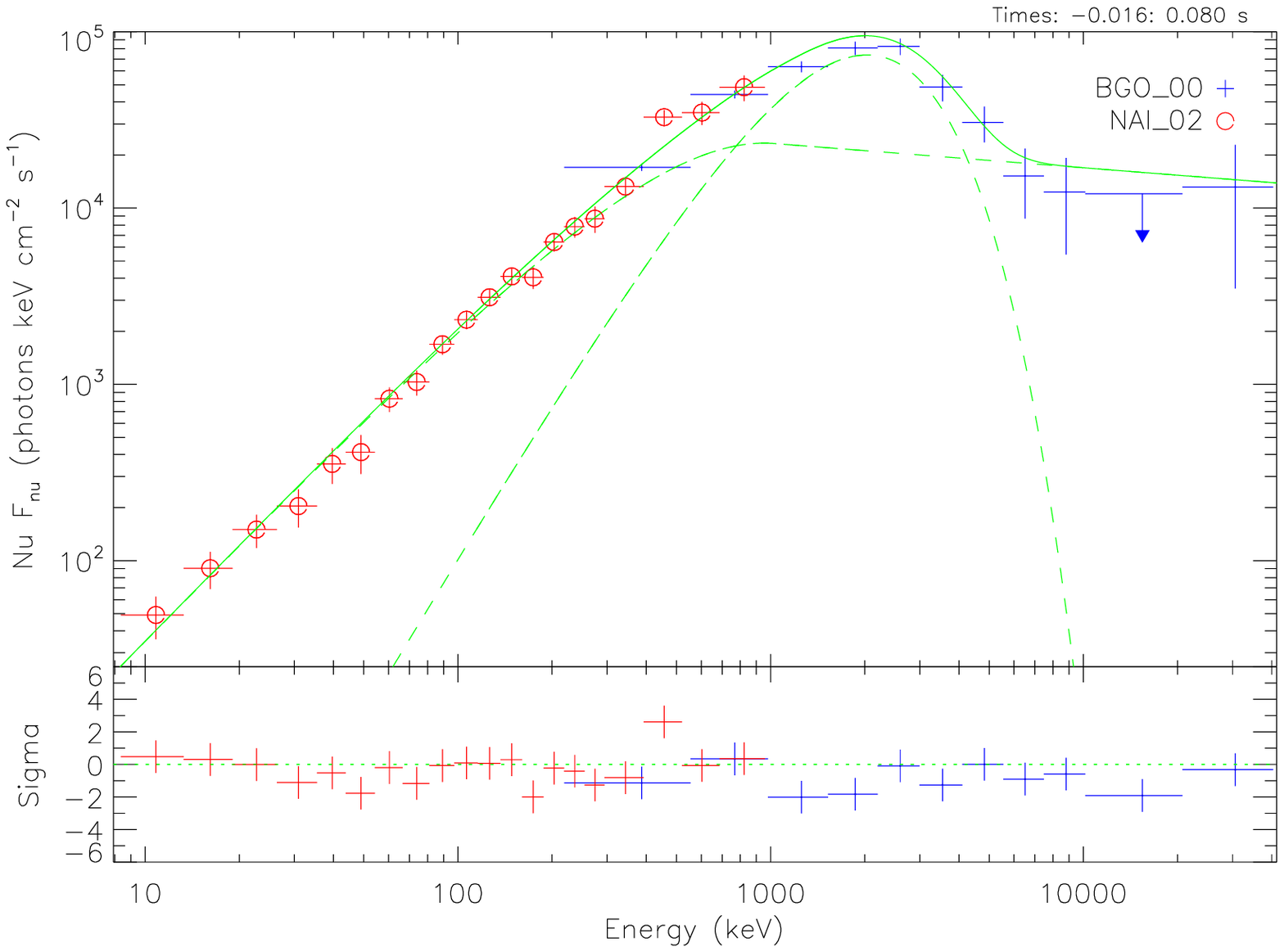} 
\hfill\null\\
\hfill
\includegraphics[width=0.42\hsize,clip]{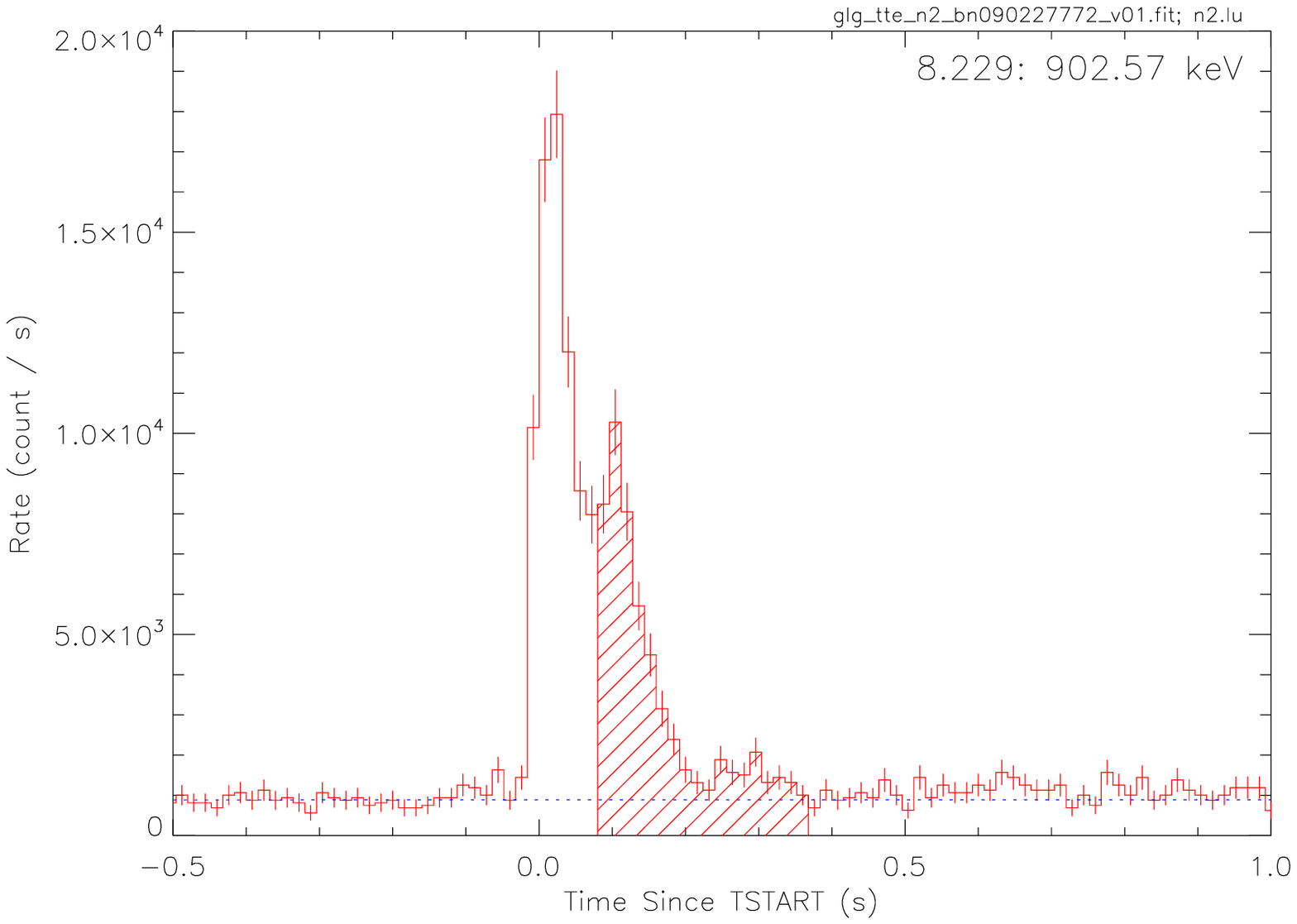}
\hfill
\includegraphics[width=0.42\hsize,clip]{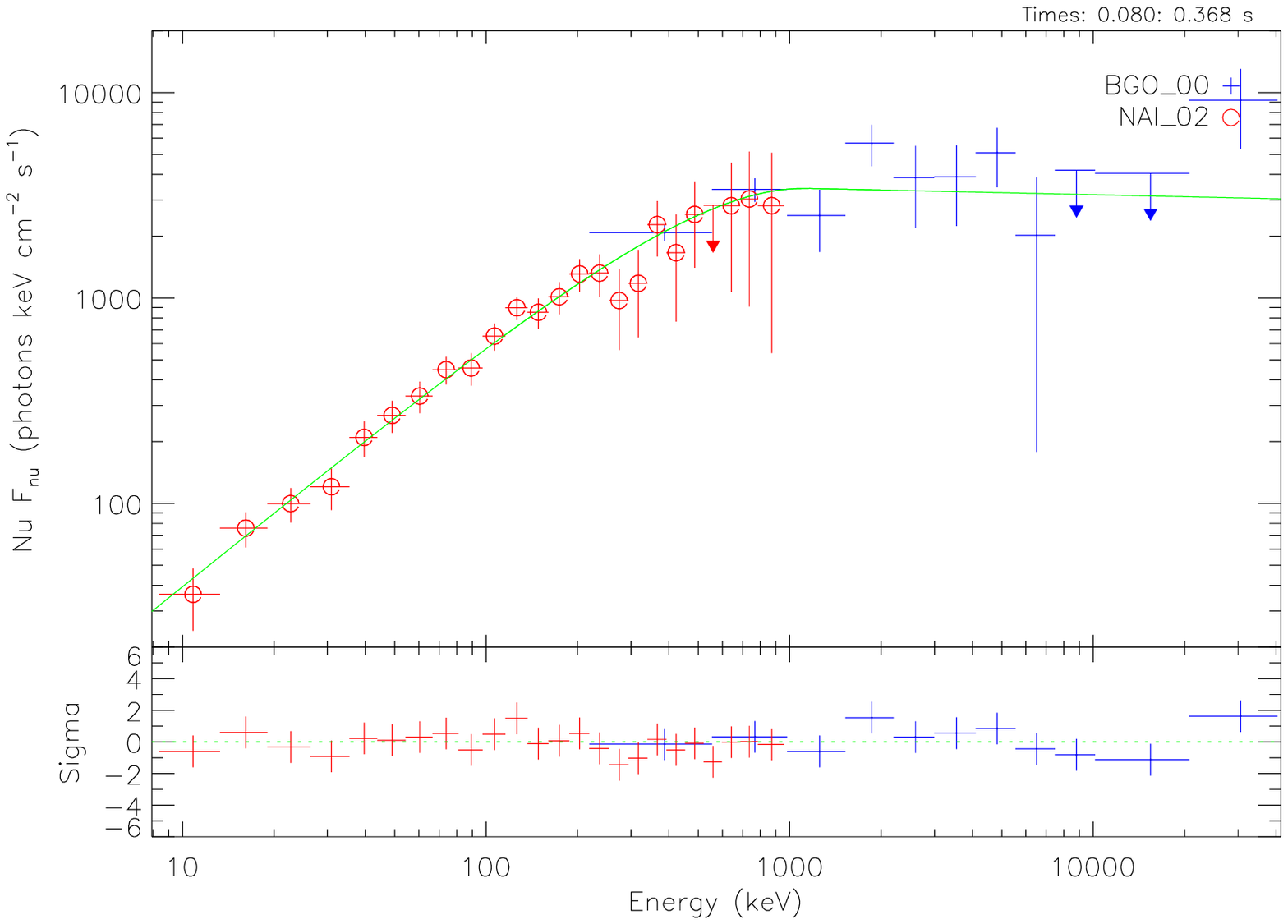} 
\hfill\null\\
\caption{The 16 ms time-binned NaI-n2 light curves of the P-GRB (left upper panel) and the extended afterglow (left lower panel) and their NaI-n2+BGO-b0 $\nu F_\nu$ spectra (on the right, the upper panel for the P-GRB and the lower one for the extended afterglow). The fit of the P-GRB is composed of a BB superimposed by a Band spectrum; the extended afterglow is well fitted by a simple Band function.}
\label{fig:2b}
\end{figure*}

We have then explored the possibility of a further rebinning of the time interval $T_{spike}$, taking advantage of the large statistical content of each time bin. 
We have plotted the NaI-n2 light curve of GRB 090227B using time bins of 16 ms (see Fig.~\ref{fig:2b}, left panels).
The re-binned light curves show two spike-like substructures. 
The duration of the first spike is $96$ ms and it is clearly distinct from the second spike.
In this time range the observed BB temperature is $kT = (517\pm 28)$ keV (see Tab.~\ref{sppgrb}) and the ratio between the fluxes of the thermal component and the non-thermal one is $F_{BB}/F_{NT} \approx 1.1$.
Consequently, we have interpreted the first spike as the P-GRB and the second spike as part of the extended afterglow.
Their spectra are shown in Fig.~\ref{fig:2b}, right panels, and the results of the spectral analysis are summarized in Tab.~\ref{sppgrb}.

\begin{table*}
\tiny
\centering
\begin{tabular}{ccccccccc}
\hline\hline
\textbf{}           &  \textbf{Model}    &  \textbf{$kT$\,[keV]}   &  \textbf{$\alpha$}  &  \textbf{$\beta$}               &  \textbf{$E_{peak}$\,[keV]}  &  \textbf{$F_{tot}$\,\textnormal{[erg/(cm${}^2$s)]}}  &  \textbf{$F_{BB}$\,\textnormal{[erg/(cm${}^2$s)]}}  &  \textbf{C-STAT/DOF } \\
\hline
\textbf{P-GRB}      &  \textbf{Band+BB}  &  $517 \pm 28$           &  $-0.80 \pm 0.05$   &  $-2.14 \pm 0.17$              &  $952 \pm 251$               &  $(3.13 \pm 0.13)\times 10^{-4}$            
&  $(1.61 \pm 0.47)\times 10^{-4}$                 &  $263.51/239$         \\
\textbf{Ext. Aft.}  &  \textbf{Band}     &                         &  $-0.79 \pm 0.06$   &  $-2.01 \pm 0.10$              &  $1048 \pm 178$              &  $(2.66 \pm 0.26)\times 10^{-5}$            
&                                                  &  $276.50/241$         \\
\hline
\end{tabular}
\caption{The results of the spectral analysis of the P-GRB (from $T_0-0.016$ s to $T_0+0.080$ s, best fit BB+Band model) and the extended afterglow (from $T_0+0.080$ s to $T_0+0.368$ s, best fit Band model) of GRB 090227B in the energy range $8$ keV -- $40$ MeV.}
\label{sppgrb}
\end{table*}

\section{Analysis of GRB 090227B in the Fireshell model}\label{sec:3}

The identification of the P-GRB is fundamental in order to determine the Baryon load and the other physical quantities characterizing the plasma at the transparency point (see Fig.~\ref{fig:4}).
Crucial is the determination of the cosmological redshift, which can be derived combining the observed fluxes and the spectral properties of the P-GRB and of the extended afterglow with the equation of motion of our theory. 
From the cosmological redshift we derive $E_{e^+e^-}^{tot}$ and the relative energetics of the P-GRB and of the extended afterglow components (see Fig.~\ref{fig:4}).
Having so derived the Baryon load $B$ and the energy $E_{e^+e^-}^{tot}$, we can constrain the total energy and simulate the canonical light curve of the GRBs with their characteristic pulses, modeled by a variable number density distribution of the CBM around the burst site.

\subsection{Estimation of the redshift of GRB 090227B}\label{sec:z}

Having determined the redshift of the source, the analysis consists of equating $E_{e^+e^-}^{tot} \equiv E_{iso}$ (namely $E_{iso}$ is a lower limit on $E_{e^+e^-}^{tot}$) and inserting a value of the Baryon load to complete the simulation.
The right set of $E_{e^+e^-}^{tot}$ and $B$ is determined when the theoretical energy and temperature of the P-GRB match the observed ones of the thermal emission [namely $E_{P-GRB} \equiv E_{BB}$ and $kT_{obs} = kT_{blue}/(1+z)$].

In the case of GRB 090227B we have estimated the ratio $E_{P-GRB}/E_{e^+e^-}^{tot}$ from the observed fluences
\begin{equation}
\label{fluences}
\frac{E_{P-GRB}}{E_{e^+e^-}^{tot}} = \frac{4\pi d_l^2 F_{BB} \Delta t_{BB}/(1+z)}{4\pi d_l^2 F_{tot} \Delta t_{tot}/(1+z)} = \frac{S_{BB}}{S_{tot}}\ ,
\end{equation}
where $d_l$ is the luminosity distance of the source and $S = F \Delta t$ are the fluences.
The fluence of the BB component of the P-GRB (see Tab~\ref{sppgrb}, first interval) is $S_{BB} = (1.54\pm0.45)\times10^{-5}$ erg/cm$^2$.
The total fluence of the burst is $S_{tot} = (3.79\pm0.20)\times10^{-5}$ erg/cm$^2$ and has been evaluated in the time interval from $T_0-0.016$ s to $T_0+0.896$ s.
This interval slightly differs from the $T_{90}$ because of the new time boundaries defined after the rebinning of the light curve at resolution of $16$ ms.
Therefore the observed energy ratio is $E_{P-GRB}/E_{e^+e^-}^{tot} = (40.67\pm0.12)\%$.
As is clear from the bottom right diagram in Fig.~\ref{fig:4}, for each value of this ratio we have a range of possible parameters $B$ and $E^{tot}_{e^+e^-}$. 
In turn, for each value of them we can determine the theoretical blue-shifted toward the observer temperature $kT_{blue}$ (see top right diagram in Fig.~\ref{fig:4}).
Correspondingly, for each couple of value of $B$ and $E^{tot}_{e^+e^-}$ we estimate the value of $z$ by the ratio between $kT_{blue}$ and the observed temperature of the P-GRB $kT_{obs}$,
\begin{equation}
\label{zth} \frac{kT_{blue}}{kT_{obs}}=1+z\ .
\end{equation}
In order to remove the degeneracy $[E_{e^+e^-}^{tot}(z),B(z)]$, we have made use of the isotropic energy formula
\begin{equation}
\label{correction}
E_{iso} = 4\pi d_l^2 \frac{S_{tot}}{(1+z)} \frac{\int^{E_{max}/(1+z)}_{E_{min}/(1+z)}{E\,N(E) dE}}{\int^{40000}_{8}{E\,N(E) dE}}\ ,
\end{equation}
in which $N(E)$ is the photon spectrum of the burst and the integrals are due to the bolometric correction on $S_{tot}$.
The correct value is the one for which the condition $E_{iso} \equiv E_{e^+e^-}^{tot}$ is satisfied. 

We have found the equality at $z = 1.61\pm0.14$ for $B = (4.13\pm0.05)\times10^{-5}$ and $E_{e^+e^-}^{tot} = (2.83\pm0.15)\times10^{53}$ ergs.
The complete quantities so determined are summarized in Tab.~\ref{par}.

\begin{table}
\centering
\begin{tabular}{cc}
\hline\hline
\textbf{Fireshell Parameter}                         & \textnormal{\textbf{Value}}     \\
\hline
$E^{tot}_{e^+e^-}$\,[erg]                            &  $(2.83 \pm 0.15)\times10^{53}$ \\
$B$                                                  &  $(4.13\pm0.05)\times10^{-5}$   \\
$\Gamma_{tr}$                                        &  $(1.44\pm0.01)\times10^4$      \\
$r_{tr}$\,[cm]                                       &  $(1.76\pm0.05)\times10^{13}$   \\
$kT_{blue}$\,[keV]                                   &  $(1.34\pm0.01)\times10^3$      \\
$z$                                                  &  $1.61 \pm 0.14$                \\
\hline
$\langle n \rangle$ [\textnormal{particles/cm}$^3$]  &  $(1.90\pm0.20)\times10^{-5}$   \\
$\langle \delta n/n \rangle$                         &  $0.82\pm0.11$                  \\
\hline
\end{tabular}
\caption{The results of the simulation of GRB 090227B in the Fireshell model.}
\label{par}
\end{table}

\subsection{The analysis of the extended afterglow and the observed spectrum of the P-GRB}\label{sec:pgrb}

As recalled in Sec.~\ref{sec:fireshell}, the arrival time separation between the P-GRB and the peak of the extended afterglow is a function of $E_{e^+e^-}^{tot}$ and $B$ and depends on the detailed profile of the CBM density.
For $B \sim 4\times10^{-5}$ (see Fig.~\ref{fig:2c}) the time separation is $\sim 10^{-3}$--$10^{-2}$ s in the source cosmological rest frame.
In this light, there is an interface between the reaching of transparency of the P-GRB and the early part of the extended afterglow. 
This connection has already been introduced in literature \citep{Peer2010,Izzo2012,Penacchioni2011}. 

\begin{table}
\centering
\begin{tabular}{ccc}
\hline\hline
\textbf{Cloud}  &  \textbf{Distance [cm]}  &  \textbf{$n_{CBM}$ [$\#$/cm$^3$]}      \\
\hline
$1^{th}$        &  $1.76\times10^{15}$     &  $(1.9\pm0.2)\times10^{-5}$            \\
$2^{th}$        &  $1.20\times10^{17}$     &  $(3.5\pm0.6)\times10^{-6}$            \\
$3^{th}$        &  $1.65\times10^{17}$     &  $(9.5\pm0.5)\times10^{-6}$            \\
$4^{th}$        &  $1.80\times10^{17}$     &  $(5.0\pm0.5)\times10^{-6}$            \\
$5^{th}$        &  $2.38\times10^{17}$     &  $(2.6\pm0.2)\times10^{-5}$            \\
$6^{th}$        &  $2.45\times10^{17}$     &  $(1.0\pm0.5)\times10^{-7}$            \\
$7^{th}$        &  $4.04\times10^{17}$     &  $(6.0\pm1.0)\times10^{-5}$            \\
\hline
\end{tabular}
\caption{The density mask of GRB 090227B: in the first column we report the number of CBM clouds, in the second one their distance away from the BH, and in the third one the number density with the associated error box.}
\label{npar}
\end{table}

\begin{figure}
\centering
\includegraphics[width=0.8\hsize,clip]{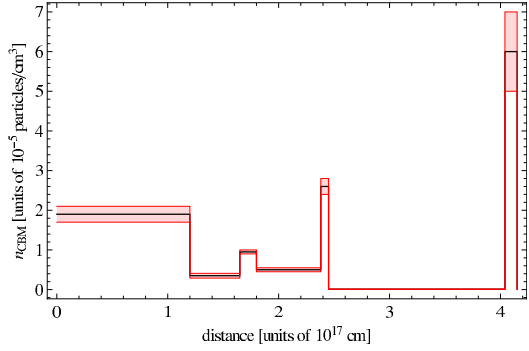} 
\caption{The radial CBM density distribution of GRB 090227B (black line) and its range of validity (red shaded region).}
\label{fig:2e}
\end{figure}

\begin{figure}
\centering
\includegraphics[height=0.8\hsize,angle=-90,clip]{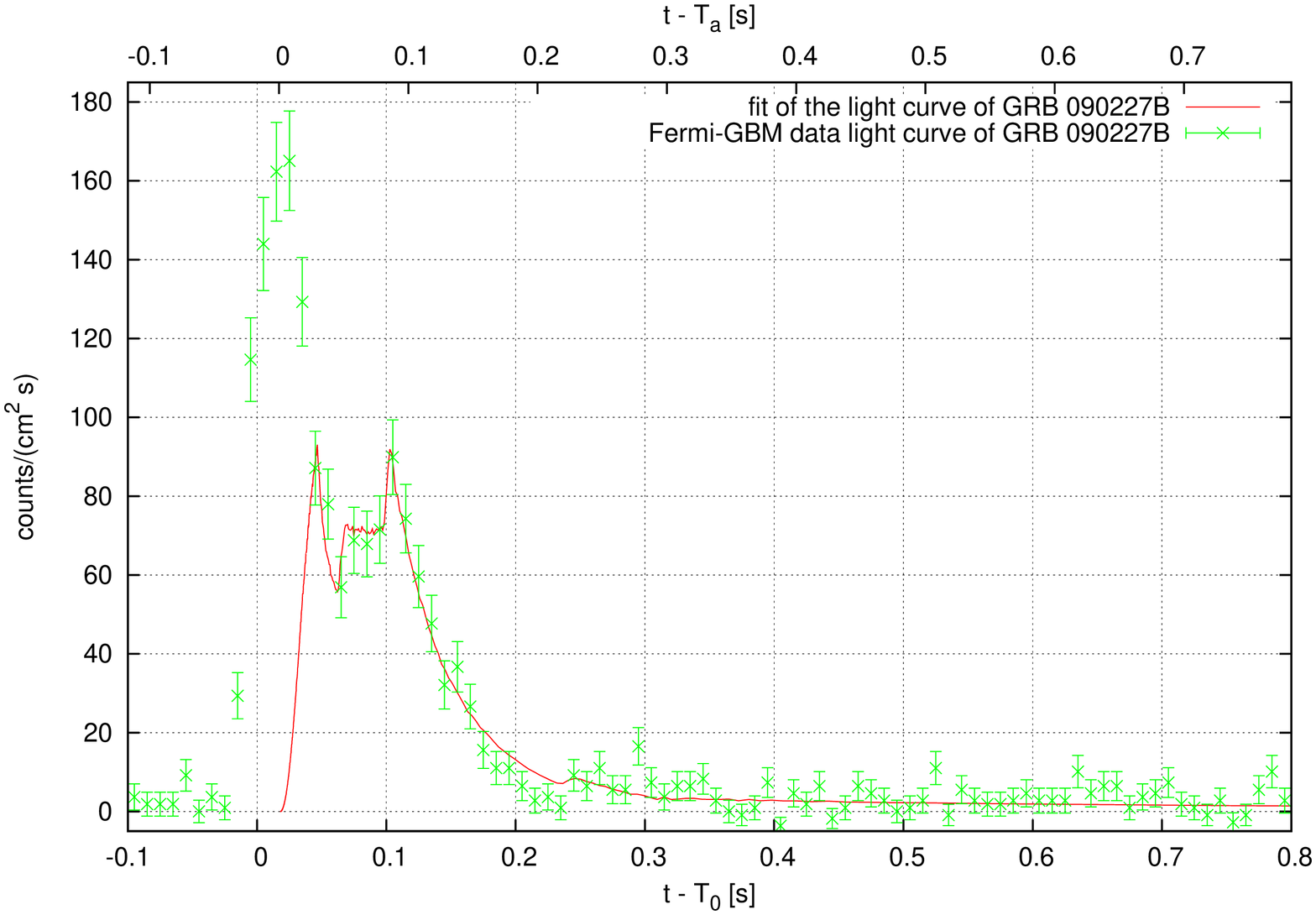}
\caption{The NaI-n2 simulated light curve of the extended-afterglow of GRB 090227B; each spike corresponds to the CBM density profile described in Tab.~\ref{npar} and Fig.~\ref{fig:2e}. The zero of the lower $x$-axis corresponds to the trigtime $T_0$; the zero of the upper $x$-axis is the time from which we have started the simulation of the extended afterglow, $T_a$, namely $0.017$ s after $T_0$.}
\label{fig:2d1}
\end{figure} 

\begin{figure*}
\centering
\includegraphics[height=0.49\hsize,angle=-90,clip]{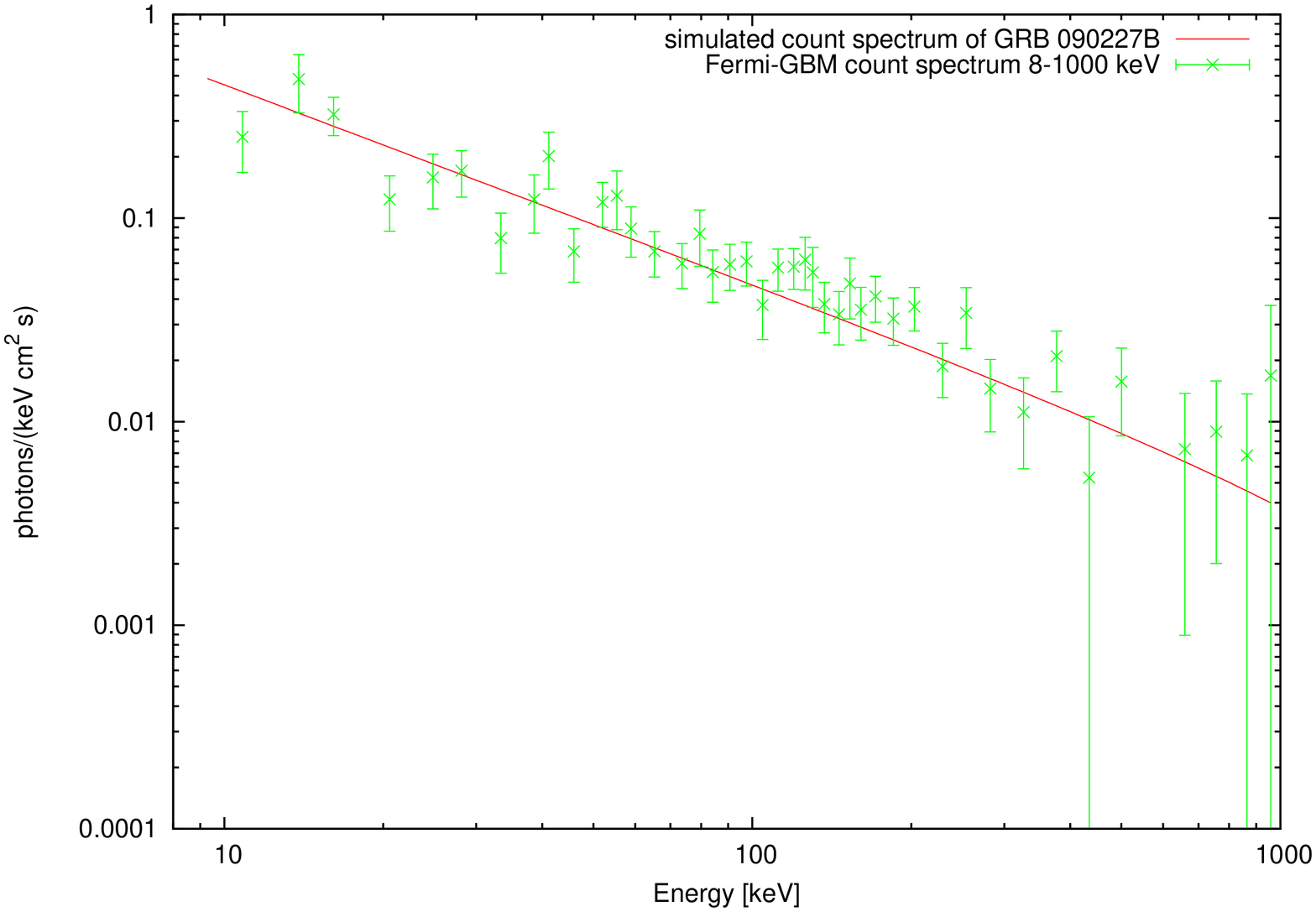}
\includegraphics[height=0.49\hsize,angle=-90,clip]{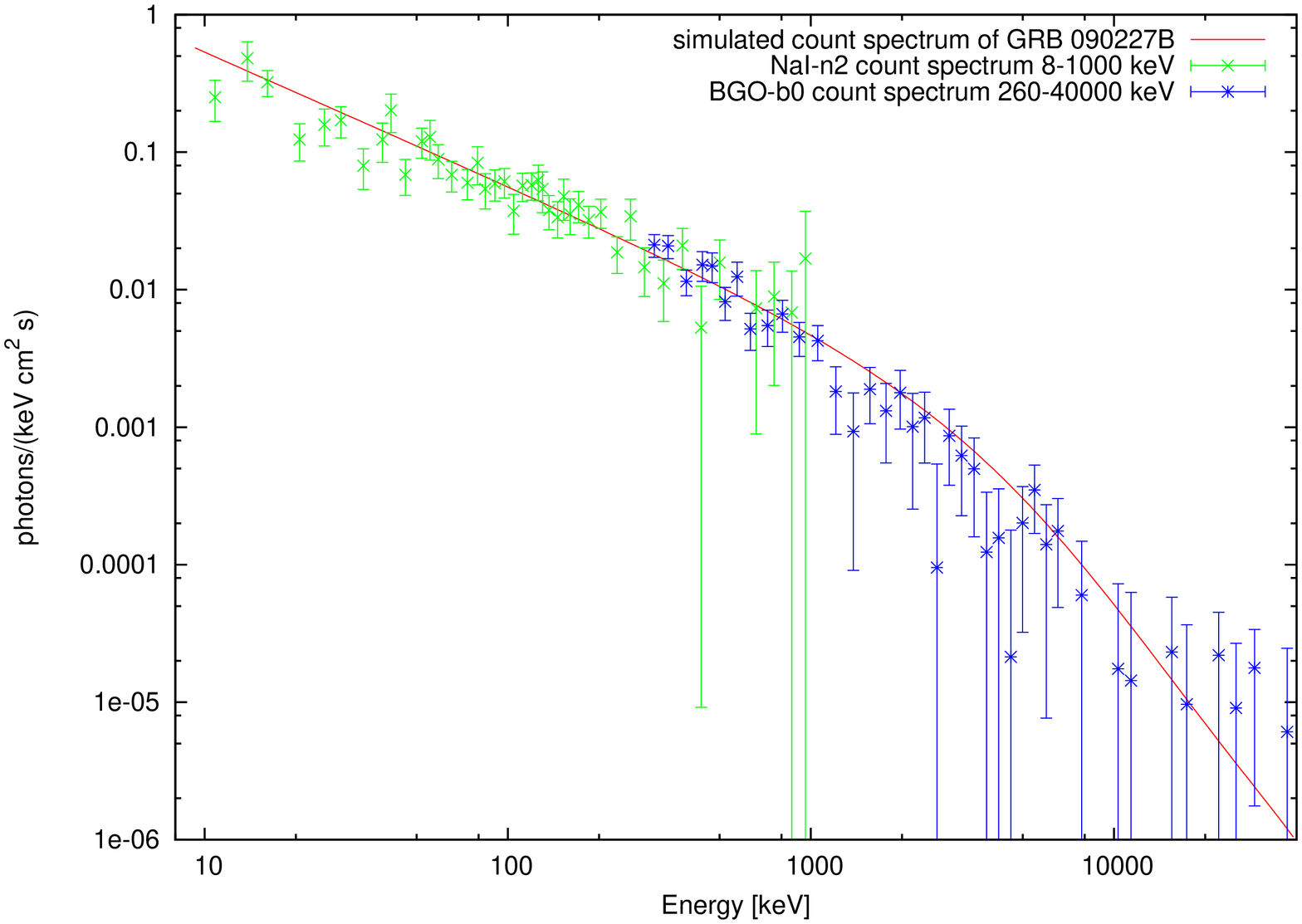}
\caption{Left panel: the simulated photon number spectrum of the extended-afterglow of GRB 090227B (from $T_0+0.015$ s to $T_0+0.385$ s) in the energy band $8$--$1000$ keV, compared to the NaI-n2 data in the same time interval. Right panel: the same simulated spectrum, with the same parameters, extended up to $40$ MeV and compared to the NaI-n2 and the BGO-b0 data in the same time interval.}
\label{fig:2d2}
\end{figure*} 

From the determination of the initial values of the energy, $E^{tot}_{e^+e^-}=2.83\times10^{53}$ ergs, of the Baryon load, $B=4.13\times10^{-5}$, and of the Lorentz factor $\Gamma_{tr}=1.44\times10^4$, we have simulated the light curve of the extended afterglow by deriving the radial distribution of the CBM clouds around the burst site (see Tab.~\ref{npar} and Fig.~\ref{fig:2e}). 
In particular, each spike in Fig.~\ref{fig:2e} corresponds to a CBM cloud.
The error boxes on the number density on each cloud is defined as the maximum possible tolerance to ensure the agreement between the simulated light curve and the observed data.
The average value of the CBM density is $\langle n \rangle = (1.90\pm0.20)\times10^{-5}$ particles/cm$^3$ with an average density contrast $\langle \delta n/n \rangle = 0.82\pm0.11$ (see also Tab.~\ref{par}).
These values are typical of the galactic halos environment.  
The filling factor varies in the range $9.1\times10^{-12} \leq \mathcal{R} \leq 1.5\times10^{-11}$, up to $2.38\times10^{17}$ cm away from the burst site, and then drops to the value $\mathcal{R} = 1.0\times10^{-15}$.
The value of the $\alpha$ parameter has been found to be $-1.99$ along the total duration of the GRB.
In Fig.~\ref{fig:2d1} we show the NaI-n2 simulated light curve ($8$--$1000$ keV) of GRB 090227B and in Fig.~\ref{fig:2d2} (left panel) the corresponding spectrum in the early $\sim 0.4$ s of the emission, using the spectral model described by Eq.~\ref{modBB} \citep{Patricelli}.
The simulation of the extended afterglow starts $T_a-T_0\sim0.017$ s after the Trigtime $T_0$.
After the submission of this manuscript, at the 13$^{th}$ \textit{Marcel Grossmann} meeting Dr. G.~Vianello suggested to extend our simulations from $1$ MeV all the way to $40$ MeV, since significant data are available from the BGO detector.
Without changing the parameters used in the theoretical simulation of the NaI-n2 data, we have extended the simulation up to $40$ MeV and we compared the results with the BGO-b0 data (see Fig.~\ref{fig:2d2}, right panel).
The theoretical simulation we performed, optimized on the NaI-n2 data alone, is perfectly consistent with the observed data all over the \emph{entire} range of energies covered by the \textit{Fermi}-GBM detector, both NaI and BGO.

\begin{figure*}
\centering
\includegraphics[width=0.49\hsize,clip]{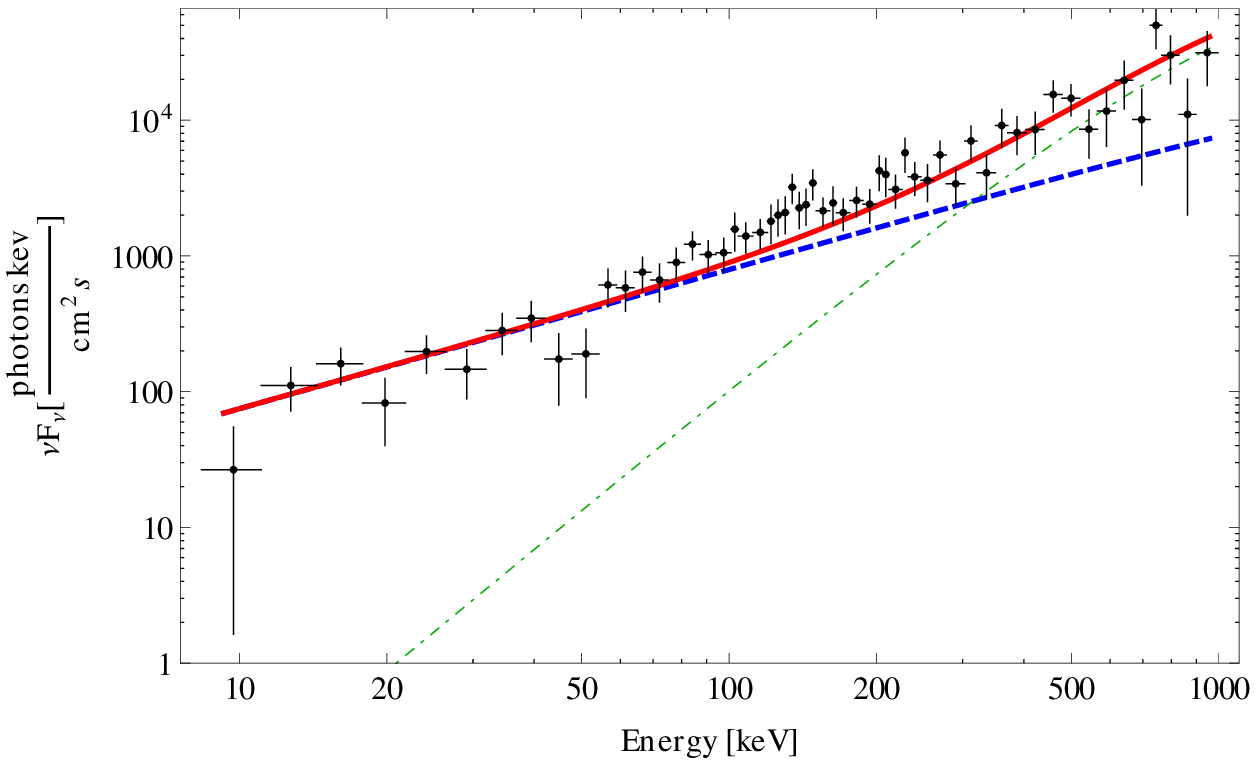}
\includegraphics[width=0.49\hsize,clip]{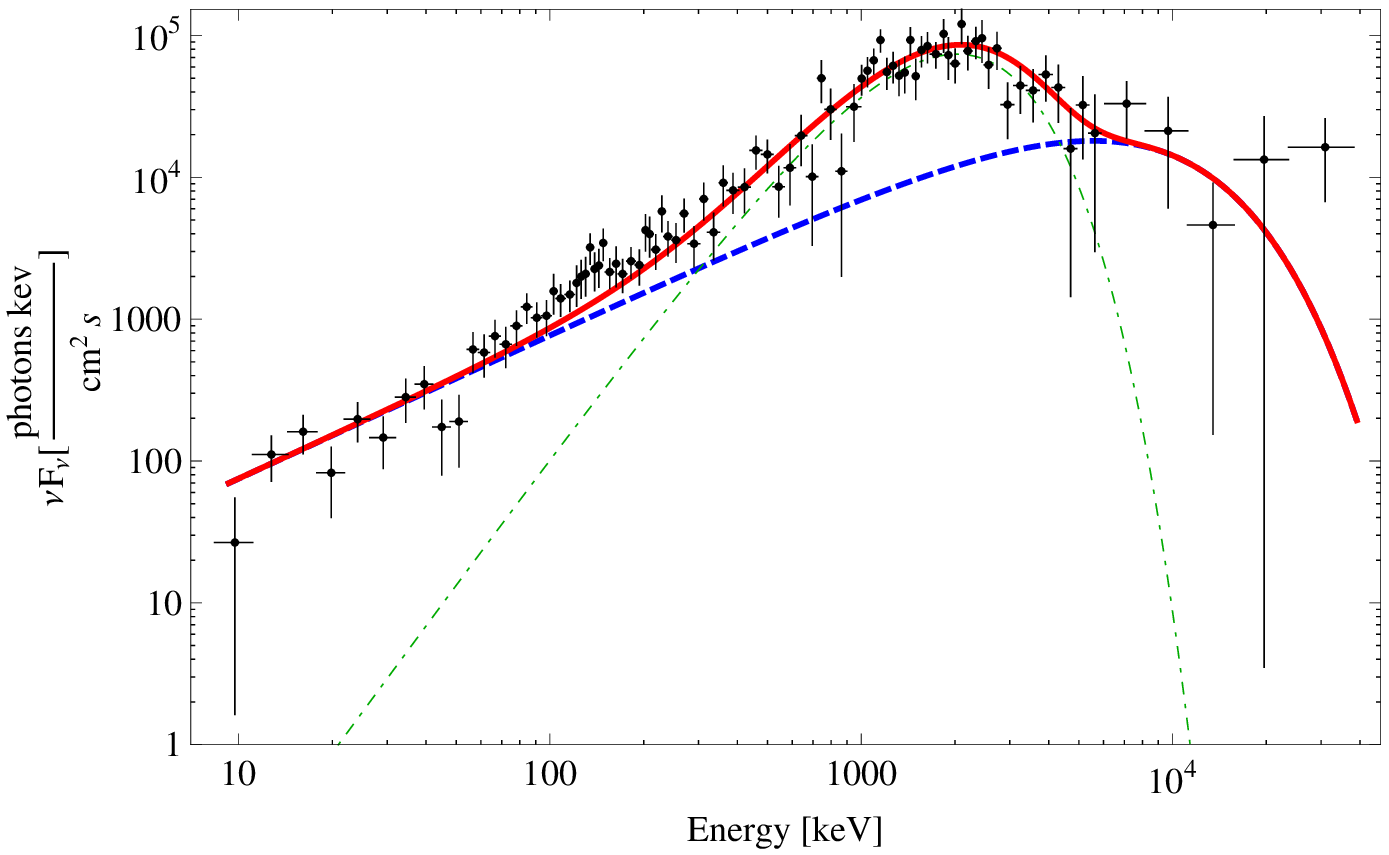}
\caption{Left panel: the time-integrated (from $T_0+0.015$ s to $T_0+0.080$ s) Fireshell simulation in the energy band $8$--$1000$ keV, dashed blue line, and the BB emission, dashed-dotted green line; the sum of the two components, the solid red line, is compared to the observed P-GRB emission. Right panel: the same considerations including the BGO data up to $40$ MeV.}
\label{fig:2f}
\end{figure*}

We turn now to the emission of the early $96$ ms.
We have studied the interface between the P-GRB emission and the on-set of the extended afterglow emission.
In Fig.~\ref{fig:2f} we have plotted the thermal spectrum of the P-GRB and the Fireshell simulation (from $T_0+0.015$ s to $T_0+0.080$ s) of the early interaction of the extended afterglow.
The sum of these two components is compared with the observed spectrum from the NaI-n2 detector in the energy range $8$--$1000$ keV (see Fig.~\ref{fig:2f}, left panel).
Then, again, from the theoretical simulation in the energy range of the NaI-n2 data, we have verified the consistency of the simulation extended up to $40$ MeV with the observed data all over the range of energies covered by the \textit{Fermi}-GBM detector, both NaI and BGO.
The result is shown in Fig.~\ref{fig:2f} (right panel).

\section{Consistency with the opacity due to pair production}\label{sec:4}

It is interesting to compare the Lorentz $\Gamma$ factor theoretically determined from the P-GRB analysis with the lower limit coming from the opacity argument applied to the afterglow emission.

An estimate on this lower limit comes from the solution of the classical compactness problem for GRBs which arises from the combination of their large energy released, $\sim 10^{51}$ ergs, the short variability time scale $\delta t$ of a few milliseconds and the observed hard non-thermal spectrum.
Using the usual (Newtonian) causality limit on the size $R \leq c\delta t$ to estimate the density of photons, one finds that the optical depth for pair production at the source $\gamma\gamma\rightarrow e^+e^-$ would be $\sim 10^{15}$ \cite[see][]{Piran99}. 
Such an optically thick source could not emit the observed non-thermal spectrum.

As \citet{Ruderman} pointed out, relativistic effects can solve this problem.
The causality limit of a source moving relativistically with Lorentz factor $\Gamma \gg 1$ towards us is $R \leq c\delta t/\Gamma^2$.
Additionally the observed photons have been blue-shifted.
At the source they have lower energy, by a factor $\approx 1/\Gamma$, which may be insufficient for pair production.
Together this leads to a decrease in the estimated optical depth by a factor $\Gamma^{2+2\beta}$ \citep{Piranmgm11}, where $\beta$ is the high energy spectral index of the photon number distribution.
Thus, the average optical depth, up to a factor due to the cosmological effects, is
\begin{equation}
\tau_{\gamma\gamma} =\frac{f_p}{\Gamma^{2+2\beta}} \frac{\sigma_T S d_l^2}{c^2 \delta t^2 m_e c^2}\ ,
\label{piran3}
\end{equation}
where $f_p$ is the fraction of photon pairs at the source that can effectively produce pairs, $\sigma_T$ is the Thompson cross-section and $S$ is the observed fluence.
From the condition $\tau_{\gamma\gamma} < 1$, Eq.~\ref{piran3} becomes 
\begin{equation}
\Gamma > \left(\frac{f_p\sigma_T S d_l^2}{c^2 \delta t^2 m_e c^2}\right)^{\frac{1}{2+2\beta}}\ .
\label{piran4}
\end{equation}
By setting $\delta t$ equal to minimum variability time scale observed for GRB 090227B, $\sim 2$ ms \citep{Guiriec2010}, and using the observed total fluence, $S_{tot} = 3.79\times10^{-5}$ erg/cm$^2$, the high energy spectral index, $\beta=2.90$, and the theoretically inferred redshift, $z = 1.61$, we obtain a lower limit $\Gamma > 594$.

The large quantitative difference between our theoretically estimated Lorentz factor from the P-GRB and the one derived from the opacity argument is not surprising in view of the very different approximations adopted.
While the determination from the P-GRB consists in a precise analysis at the instant of transparency, the determination of the lower limit from the  Eq.~\ref{piran4} is based on an estimate taking a time-averaged value on the entire extended afterglow.

Important, of course, is that the precise value determined from the P-GRB does fulfill the inequality given in Eq.~\ref{piran4}.

\section{Conclusions}\label{sec:5}

The comprehension of this short GRB has been improved by analyzing the different spectra in the $T_{90}$, $T_{spike}$ and $T_{tail}$ time intervals. 
We have shown that within the $T_{90}$ and the $T_{spike}$ all the considered models (BB+Band, Band+PL, Compt+PL) are viable, while in the $T_{tail}$ time interval, the presence of a thermal component is ruled out.
The result of the analysis in the $T_{tail}$ time interval gives an additional correspondence between the Fireshell model (see Sec.~\ref{sec:extaft}) and the observational data. 
According to this picture, the emission within the $T_{spike}$ time interval is related to the P-GRB and it is expected to have a thermal spectrum with in addition an extra NT component due to an early onset of the extended afterglow. 
In this time interval a BB with an additional Band component has been observed and we have shown that it is statistically equivalent to the Compt+PL and the Band+PL models.
Our theoretical interpretation is consistent with the observational data and statistical analysis.
From an astrophysical point of view the BB+Band model is preferred over the Compt+PL and the Band+PL models, been described by a consistent theoretical model.

GRB 090227B is the missing link between the genuine short GRBs, with the Baryon load $B \lesssim 5\times10^{-5}$ and theoretically predicted by the Fireshell model \citep{Ruffini2001c,Ruffini2001,Ruffini2001a}, and the long bursts.

From the observations, GRB 090227B has an overall emission lasting $\sim 0.9$ s with a fluence of $3.79 \times 10^{-5}$ erg/cm$^2$ in the energy range $8$ keV -- $40$ MeV.
In absence of an optical identification, no determination of its cosmological redshift and of its energetics was possible.

Thanks to the excellent data available from Fermi-GBM \citep{Meegan2009}, it has been possible to probe the comparison between the observation and the theoretical model.
In this sense, we have then performed a more detailed spectral analysis on the time scale as short as $16$ ms of the time interval $T_{spike}$.
As a result we have found in the early $96$ ms a thermal emission which we have identified with the theoretically expected P-GRB component.
The subsequent emission of the time interval $T_{spike}$ has been interpreted as part the extended afterglow.
Consequently, we have determined the cosmological redshift, $z = 1.61\pm0.14$, as well as the Baryon load, $B = (4.13\pm0.05)\times10^{-5}$, its energetics, $E^{tot}_{e^+e^-} = (2.83\pm0.15)\times10^{53}$ ergs, and the extremely high Lorentz $\Gamma$ factor at the transparency, $\Gamma_{tr} = (1.44\pm0.01)\times10^4$.

We are led to the conclusion \citep[see also][]{Rueda2012} that the progenitor of this GRB is a binary neutron star, which for simplicity we assume to have the same mass, by the following considerations:
\begin{enumerate}
\item the very low average number density of the CBM, $\langle n_{CBM}\rangle \sim 10^{-5}$ particles/cm$^3$; this fact points to two compact objects in a binary system that have spiraled out in the halo of their host galaxy \cite[see][]{Bernardini2007,Bernardini2008,Bianco2008,Caito2009,Caito2010,deBarros2011};
\item the large total energy, $E^{tot}_{e^+e^-} = 2.83\times10^{53}$ ergs, which we can indeed infer in view of the absence of beaming, and the very short time scale of emission point again to two neutron stars. We are led to a binary neutron star with total mass $m_1+m_2$ larger than the neutron star critical mass, $M_{cr}$. In light of the recent neutron star theory in which all the fundamental interactions are taken into account \citep{Belvedere}, we obtain for simplicity in the case of equal neutron star masses, $m_1 = m_2 = 1.34M_\odot$, radii $R_1 = R_2 = 12.24$ km, where we have used the NL3 nuclear model parameters for which $M_{cr}=2.67M_\odot$;
\item the very small value of the Baryon load, $B = 4.13\times10^{-5}$, is consistent with the above two neutron stars which have crusts $\sim 0.47$ km thick. The new theory of the neutron stars, developed in \citet{Belvedere}, leads to the prediction of GRBs with still smaller Baryon load and, consequently, shorter periods. We indeed infer an absolute upper limit on the energy emitted via gravitational waves, $\sim 9.6\times10^{52}$ ergs \citep[see][]{Rueda2012}.
\end{enumerate}

\begin{figure}
\centering
\includegraphics[width=\hsize,clip]{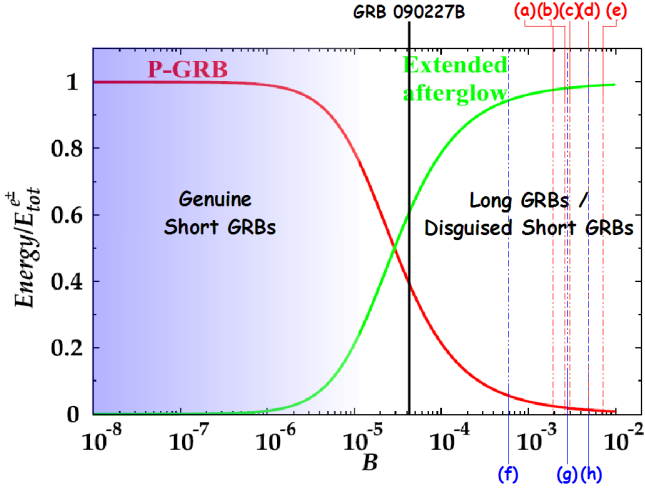}
\caption{The energy emitted in the extended afterglow (green curve) and in the P-GRB (red curve) in units of the total energy $E_{e^+e^-}^{tot} = 1.77 \times 10^{53}$ erg are plotted as functions of the B parameter. In the figure are also marked some values of the Baryon load: in black GRB 090227B and in red and blue some values corresponding to, respectively, some long and disguised short GRBs we analyzed.}
\label{fig:s}
\end{figure}

\begin{table}
\tiny
\centering
\begin{tabular}{ccccc}
\hline\hline
\textbf{label} & \textbf{GRB} & $E^{tot}_{e^+e^-}$ [erg] & $B$ & $\langle n_{CBM}\rangle$ [\#/cm$^3$] \\
\hline
(a) & 090618  & $2.49\times10^{53}$ & $1.98\times10^{-3}$ & $1.0$ \\
(b) & 080319B & $1.32\times10^{54}$ & $2.50\times10^{-3}$ & $6.0$ \\
(c) & 991216  & $4.83\times10^{53}$ & $3.00\times10^{-3}$ & $1.0$ \\
(d) & 030329  & $2.12\times10^{52}$ & $4.80\times10^{-3}$ & $2.0$ \\
(e) & 031203  & $1.85\times10^{50}$ & $7.40\times10^{-3}$ & $0.3$ \\
\hline
(f) & 050509B & $5.52\times10^{48}$ & $6.00\times10^{-4}$ & $1.0\times10^{-3}$ \\
(g) & 060614  & $2.94\times10^{51}$ & $2.80\times10^{-3}$ & $1.0\times10^{-3}$ \\
(h) & 970228  & $1.45\times10^{54}$ & $5.00\times10^{-3}$ & $9.5\times10^{-4}$ \\
\hline
    & 090227B & $2.83\times10^{53}$ & $4.13\times10^{-5}$ & $1.9\times10^{-5}$ \\
\hline
\end{tabular}
\caption{List of the long and disguised short GRBs labeled in Fig.~\ref{fig:s} with in addition GRB 090227B. For each burst the total energy of the plasma, the Baryon load, and the average CBM density are indicated.}
\label{pars}
\end{table}

We can then generally conclude on the existence of three different possible structures of the canonical GRBs (see Fig.~\ref{fig:s} and Tab.~\ref{pars}):
\begin{enumerate}
\item[a.] long GRB with Baryon load $3.0\times10^{-4} \lesssim B \leq 10^{-2}$, exploding in a CBM with average density of $\langle n_{CBM} \rangle \approx 1$ particle/cm$^3$, typical of the inner galactic regions;
\item[b.] disguised short GRBs with the same Baryon load as the previous class, but occurring in a CBM with $\langle n_{CBM} \rangle \approx 10^{-3}$ particle/cm$^3$, typical of galactic halos \citep{Bernardini2007,Bernardini2008,Bianco2008,Caito2009,Caito2010,deBarros2011};
\item[c.] genuine short GRBs which occur for $B \lesssim 10^{-5}$ with the P-GRB predominant with respect to the extended afterglow and exploding in a CBM with $\langle n_{CBM} \rangle \approx 10^{-5}$ particle/cm$^3$, typical again of galactic halos, being GRB 090227B the first example.
\end{enumerate}
Both classes of GRBs occurring in galactic halos originate from binary mergers.

Finally, if we turn to the theoretical model within a general relativistic description of the gravitational collapse to a $10 M_\odot$ black hole, see e.g. \citet{Separatrix,Short2005} and Fig.~2 in \citet{Short2006}, we find the necessity of time resolutions of the order the fraction of a ms, possibly down to $\mu$s, in order to follow such a process. 
One would need new space missions larger collecting area to prove with great accuracy the identification of a thermal component.
It is likely that an improved data acquisition with high signal to noise on shorter time scale would evidence more clearly the thermal component as well as distinguish more effectively different fitting procedure.

\acknowledgements

We are grateful to the anonymous referee for her/his important remarks which have improved the presentation of our paper.
We thank also Dr. Giacomo Vianello for the important suggestion of checking the extrapolation from $1$ MeV up to $40$ MeV of the simulated spectra, comparing them with the Fermi-BGO data: this has provided a further important check of the consistency of our theoretical model with the data all over the range of energies covered by the \textit{Fermi}-GBM detector, both NaI and BGO.
MM is especially grateful to Jorge A. Rueda and Gregory Vereshchagin for fruitful discussions about this work.

\end{document}